# Phosphine as a Biosignature Gas in Exoplanet Atmospheres

C. Sousa-Silva[1], S. Seager[1,2,3], S. Ranjan[1], J. J. Petkowski[1], Z. Zhan[1], R. Hu[4,5], W. Bains[6]
[1] Dept. of Earth, Atmospheric, and Planetary Sciences, MIT,
77 Massachusetts Ave, Cambridge, MA 02139, USA
[2] Dept. of Physics, MIT,
77 Massachusetts Ave, Cambridge, MA 02139, USA
[3] Dept. of Aeronautics and Astronautics, MIT,
77 Massachusetts Ave, Cambridge, MA 02139, USA
[4] Jet Propulsion Laboratory, California Institute of Technology, Pasadena, CA 91109
[5] Division of Geological and Planetary Sciences, California Institute of Technology, Pasadena, CA 91125
[6] Rufus Scientific, 37 The Moor, Melbourn, Royston, Herts SG8 6ED, UK.

## Abstract

A long-term goal of exoplanet studies is the identification and detection of biosignature gases. Beyond the most discussed biosignature gas $O_2$, only a handful of gases have been considered in detail. Here we evaluate phosphine ($PH_3$). On Earth, $PH_3$ is associated with anaerobic ecosystems, and as such it is a potential biosignature gas in anoxic exoplanets.

We simulate the atmospheres of habitable terrestrial planets with $CO_2$- and $H_2$-dominated atmospheres, and find that phosphine can accumulate to detectable concentrations on planets with surface production fluxes of $10^{10}$-$10^{14}$ cm$^{-2}$ s$^{-1}$ (corresponding to surface concentrations of 10s of ppb to 100s of ppm), depending on atmospheric composition, and UV irradiation. While high, the surface flux values are comparable to the global terrestrial production rate of methane, or $CH_4$ ($10^{11}$ cm$^{-2}$ s$^{-1}$) and below the maximum local terrestrial $PH_3$ production rate ($10^{14}$ cm$^{-2}$ s$^{-1}$). As with other gases, $PH_3$ can more readily accumulate on low-UV planets, e.g. planets orbiting quiet M-dwarfs or with a photochemically generated UV shield.

If detected, phosphine is a promising biosignature gas, as it has no known abiotic false positives on terrestrial planets from any source that could generate the high fluxes required for detection. $PH_3$ also has three strong spectral features such that in any atmosphere scenario one of the three will be unique compared to other dominant spectroscopic molecules. $PH_3$'s weakness as a biosignature gas is its high reactivity, requiring high outgassing rates for detectability. We calculate that tens of hours of JWST time are required for a potential detection of $PH_3$. Yet, because $PH_3$ is spectrally active in the same wavelength regions as other atmospherically important molecules (such as $H_2O$ and $CH_4$), searches for $PH_3$ can be carried out at no additional observational cost to searches for other molecular species relevant to characterizing exoplanet habitability.

# 1. Introduction

Life makes use of thousands of volatile molecular species that could contribute towards a biosphere and its associated atmospheric spectrum. Some of these volatiles may accumulate in a planetary atmosphere and be remotely detectable; these are commonly called “biosignature gases”. Theoretical studies of biosignature gases have been recently heavily reviewed elsewhere (Grenfell 2018; Kiang *et al.* 2018; Schwieterman *et al.* 2018; Seager *et al.* 2016).

Prominent biosignature gases on Earth are those that are both relatively abundant and spectroscopically active (primarily $O_2$ and its photochemical byproduct $O_3$, but also $CH_4$ and $N_2O$). Other gases that are not prominent in Earth’s atmosphere but might be prominent in exoplanet atmospheres have also been studied, for example DMS, DMDS and $CH_3Cl$ (Domagal-Goldman *et al.* 2011; Pilcher 2003; Segura *et al.* 2005). The next generation telescopes will open the era of the study of rocky exoplanet atmospheres. The James Webb Space Telescope (JWST, planned for launch in 2021) is the most capable for transmission spectra studies of a handful of the most suitable rocky planets transiting bright M-dwarf stars (Gardner *et al.* 2006), while ESA’s Atmospheric Remote-sensing Infrared Exoplanet Large-survey (ARIEL, planned for launch in 2028) may be able to detect atmospheric components on super-Earths around the smallest M dwarf stars (Pascale *et al.* 2018). Large ground-based telescopes now under construction, i.e. GMT, ELT and TMT (Johns *et al.* 2012; Skidmore *et al.* 2015; Tamai and Spyromilio 2014), can also reach M-dwarf star rocky planets by direct imaging, with the right instrumentation.

To our knowledge phosphine ($PH_3$) has not yet been evaluated as a biosignature gas. In the Earth’s atmosphere $PH_3$ is a trace gas. It is possible, however, that biospheres on other planets could accumulate significant, detectable $PH_3$ levels. In particular, anoxic biospheres where life would not be heavily dependent on oxygen could produce $PH_3$ in significantly higher quantities than on Earth (Bains *et al.* 2019b).

Astronomical observations find that phosphine is spectroscopically active and present in stellar atmospheres (namely carbon stars) and in the giant planet atmospheres of Jupiter and Saturn (Agúndez *et al.* 2014; Bregman *et al.* 1975; Tarrago *et al.* 1992). In T-dwarfs and giant planets, $PH_3$ is expected to contain the entirety of the atmospheres’ phosphorus in the deep atmosphere layers (Visscher *et al.* 2006), where it is sufficiently hot for $PH_3$ formation to be thermodynamically favored. In both Jupiter and Saturn, $PH_3$ is found on the high observable layers at abundances (4.8 ppm and 15.9 ppm, respectively) several orders of magnitude higher than those predicted by thermodynamic equilibrium (Fletcher *et al.* 2009). This overabundance of $PH_3$ occurs because chemical equilibrium timescales are long when compared to convective timescales (Noll and Marley 1997). $PH_3$ forms in the hotter deep layers of the atmosphere (temperatures $\gtrsim$ 800 K) and is mixed upwards, so that the $PH_3$ inventory at the cloud-top is replenished. In every astronomical body, apart for Earth, where phosphine has been detected thus far, there are regions with high enough temperatures for $PH_3$ to be the thermodynamically favored phosphorus species. It has been postulated that elemental phosphorus species originating from the photolysis of $PH_3$ are responsible for the red coloring of Jupiter's red spot and other Jovian chromophores (Prinn and Lewis 1975), though this hypothesis has not achieved wide community acceptance (e.g., (Kim 1996; Noy *et al.* 1981)). For a review of chemical species that are current candidates for the chromophores of Jupiter see (Carlson *et al.* 2016) and references therein. $PH_3$ has not been detected in the observable layers of ice giants, such as Uranus and Neptune (Burgdorf *et al.* 2004; Moreno *et al.* 2009), despite these planets having sufficiently hot layers to produce $PH_3$ and strong convection currents which could transport $PH_3$ to observable altitudes. Observations put the P/H abundance in Uranus and Neptune at an upper limit of <0.1 solar P/H, which is significantly lower than expected (Teanby *et al.* 2019).

In this work, we critically assess phosphine as a biosignature gas. We first summarize in what circumstances phosphine is generated by life on Earth (Section 2.1 and Section 2.2). We next review the known destruction mechanisms for $PH_3$ (Section 2.3) and describe our inputs and methods for the assessment of the detectability of $PH_3$ in a variety of atmosphere types (Section 3). We then present our results (Section 4): here, we first calculate surface fluxes and associated atmospheric abundances required for the remote spectroscopic detection of $PH_3$ in transmission and emission spectra (Section 4.1). We then highlight the properties of the $PH_3$ spectrum that help distinguish it from other molecules (Section 4.2). Next we present thermodynamic calculations that show $PH_3$, in temperate planets, has no substantial false positives as a biosignature gas (Section 4.3). We conclude with a discussion of our results (Section 5).

## 2. Phosphine Sources and Sinks

On Earth, phosphine is associated with biological production in anaerobic environments and anthropogenic production via a multitude of industrial processes. $PH_3$ has low mean production rates on Earth but it is a mobile gas and is found globally, albeit in trace amounts, in the atmosphere. Below we summarize the known emissions of $PH_3$ on Earth (Section 2.1), $PH_3$'s association with life (Section 2.2), and its known destruction mechanisms (Section 2.3).

### 2.1 Phosphine Emissions on Earth

Phosphine is a ubiquitous trace component of the atmosphere on modern Earth (Morton and Edwards 2005). About 10% of the phosphorus in the atmosphere is $PH_3$; the major phosphorus form is phosphate, mostly as phosphoric acid (Elm *et al.* 2017). Although $PH_3$ is found everywhere in the Earth's atmosphere, its atmospheric abundance is widely variant, with high concentration regions sometimes having more $PH_3$ than low concentration areas by a factor of 10,000 (Pasek *et al.* 2014). $PH_3$ has been found worldwide in the lower troposphere of the Earth in the ppq to ppb range in daytime, with higher night time concentrations than at day time (due to inhibited UV-induced oxidation) (Gassmann 1994; Gassmann *et al.* 1996; Glindemann *et al.* 2003; Glindemann *et al.* 1996b; Han *et al.* 2000; Hong *et al.* 2010a; Ji-ang *et al.* 1999; Li *et al.* 2009; Zhang *et al.* 2010; Zhu *et al.* 2007a; Zhu *et al.* 2007b; Zhu *et al.* 2006a). In the high troposphere $PH_3$ was found at a peak of 7 ppt during daylight (Glindemann *et al.* 2003; Han *et al.* 2011b). This implies that sunlight does not lead to complete destruction of $PH_3$, unlike previous suggestions (Glindemann *et al.* 2003; Han *et al.* 2011b). A sample of locally measured gaseous $PH_3$ concentrations in a variety of environments on Earth, ranging from ppq to ppb (ng/m$^3$ to μg/m$^3$), can be found in Figure 1.

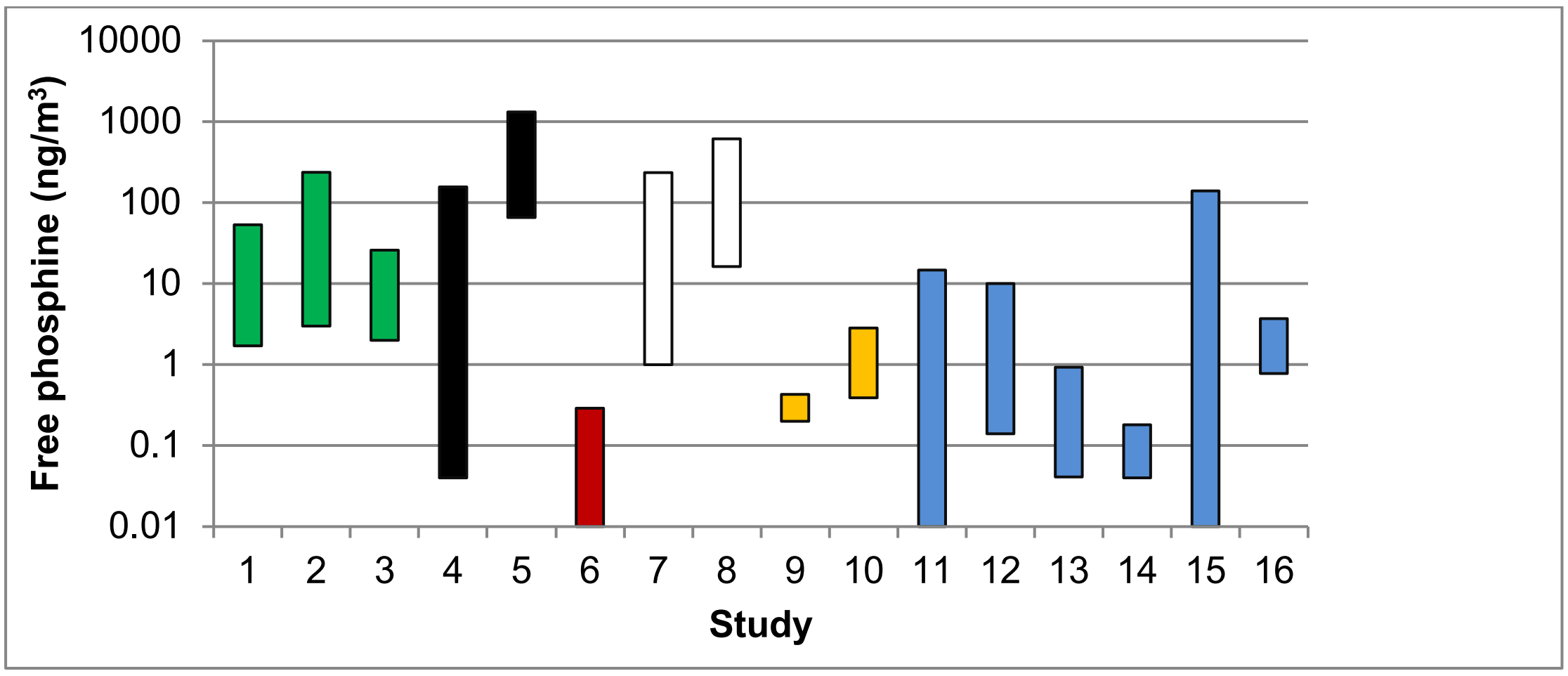

**Figure 1**. Measurements of phosphine concentrations in the Earth's atmosphere. Study number shown in x-axis (references below) and y-axis showing the span of locally measured concentration of gaseous $PH_3$ in units of ng/m$^3$, with maximum values of 600.2 and 1259 ng/m$^3$ (corresponding to concentrations ranging between ppq to ppb). Green bars: marshlands and paddy fields. Black bars – industrial environments. Red bar – Namibia (rural environment). White bars - arctic and Antarctic environments. Yellow bars - Upper troposphere. Blue bars - oceanic samples (coastal and open ocean. References for studies shown: 1) (Han *et al.* 2011a) 2) (Han *et al.* 2000) 3) (Niu *et al.* 2013) 4) (Glindemann *et al.* 1996a) 5) (Zhang *et al.* 2010) 6) (Glindemann *et al.* 1996a) 7) (Zhu *et al.* 2007a; Zhu *et al.* 2007b) 8) (Zhang *et al.* 2010) 9,10) (Glindemann *et al.* 2003) 11) (Li *et al.* 2009) 12) (Zhu *et al.* 2007a; Zhu *et al.* 2007b) 13) (Gassmann *et al.* 1996) 14) (Glindemann *et al.* 2003) 15) (Geng *et al.* 2005; Han *et al.* 2011b) 16) (Hong *et al.* 2010a). We do not include measurements of "Matrix-Bound Phosphine" (material that releases $PH_3$ when a matrix is treated with high temperatures and strong acid or alkali). Figure adapted from (Bains *et al.* 2019a). Please see (Bains *et al.* 2019a) for more details on MBP and environmental $PH_3$ production. Gaseous $PH_3$ is found in multiple altitudes in the Earth's atmosphere above a wide variety of environments, in concentrations ranging from ppq to ppb.

On Earth, a significant source of phosphine emissions is anthropogenic activity. Because of its broad toxicity to aerobic organisms[1], $PH_3$ is widely used in the agricultural industry as a rodenticide and insecticide (Bingham 2001; Chen *et al.* 2017; Devai *et al.* 1988; Glindemann *et al.* 2005; Perkins *et al.*

[1] Exposure to phosphine abundances of 400ppm results in a quick death (Fluck 1973).

2015). $PH_3$ is also used commercially, e.g., as a doping agent (Budavari *et al.* 1996). However, $PH_3$ emissions linked to biological activity are believed to form the majority of atmospheric $PH_3$ (Glindemann *et al.* 2005; Morton and Edwards 2005). Evidence for the association of $PH_3$ with anaerobic biology is presented in Section 2.2.

## 2.2 Biological Production of Phosphine

All life on Earth relies on phosphorous compounds. The biological phosphorus cycle is heavily, but not exclusively, reliant on phosphates. Other, less oxidized, phosphorus-containing molecules also play a crucial role in the phosphorus cycle (see Appendix E). The exact role of phosphine in this global phosphorus cycle is not yet fully known. It is, however, likely that, similarly to other reduced phosphorus species, $PH_3$ also has an important role in the global cycling of this essential element. Biological $PH_3$ production is associated with microbial activity in environments that are strictly anoxic (lacking oxygen). This finding is with alignment with the fact that the toxicity of $PH_3$ is intrinsically linked to its interference with $O_2$-dependent metabolism (Bains *et al.* 2019b). The majority of reports of biological $PH_3$ come from the studies of environments with anaerobic niches such as wetlands and sludges (Devai and Delaune 1995; Devai *et al.* 1988; Eismann *et al.* 1997b; Glindemann *et al.* 1996b; Roels and Verstraete 2004) and animal intestinal tracts, flatus and feces[2] (Chughtai and Pridham 1998; Eismann *et al.* 1997a; Gassmann and Glindemann 1993; Zhu *et al.* 2006a; Zhu *et al.* 2006b; Zhu *et al.* 2014).

The argument that $PH_3$ is associated with anaerobic life is strengthened by its detection in a wide variety of ecosystems with anoxic niches, including above penguin colonies, rich in bird guano, where it reaches abundances of 300 ppt (Hong *et al.* 2010a; Ji-ang *et al.* 1999; Li *et al.* 2009; Zhu *et al.* 2007a; Zhu *et al.* 2006a); paddy fields[3] (Chen *et al.* 2017; Han *et al.* 2011a); rivers and lakes (Ding *et al.* 2014; Feng *et al.* 2008; Geng *et al.* 2010; Han *et al.* 2011b; Hong *et al.* 2010b); marshlands (Devai and Delaune 1995; Eismann *et al.* 1997b); and landfills and sludges (Ding *et al.* 2005a; Ding *et al.* 2005b; Roels and Verstraete 2004). Several studies have also reported the production of $PH_3$ from mixed bacterial cultures in the lab (Jenkins *et al.* 2000; Liu *et al.* 2008; Rutishauser and Bachofen 1999), in one case bacteria turning half the phosphorus in the culture medium (~180mg/L) into $PH_3$ in 56 days (Devai *et al.* 1988).

Despite a large body of robust circumstantial evidence for the production of phosphine by life, the exact mechanisms for biologically-associated production of $PH_3$ are still debated, and the metabolic pathway leading to $PH_3$ production in anaerobic bacteria is unknown. However, we note that the absence of a known enzymatic mechanism is not evidence for the absence of biological production. The synthetic pathways for most of life's natural products are not known, and yet their origin is widely accepted to be biological because of the implausibility of their abiotic synthesis, their obligate association with life, and their chemical similarity to other biological products. For example, a recently published, manually curated, database of natural molecules produced by life on Earth contains ~220,000 unique molecules of biological origin, produced by thousands of species (Petkowski *et al.* 2019a) while the number of known, experimentally elucidated, metabolic pathways from organisms belonging to all three domains of life is only ~2,720 (Caspi *et al.* 2017). Further examples of the complexities in discovering metabolic pathways for molecules associated with biological activity are provided in Appendix E.

[2] We note that animal guts are anaerobic, even in small animals.
[3] In agricultural wetlands both industrial and biological sources of $PH_3$ are plausible.

There are two proposed explanations for the production of $PH_3$ in anoxic ecosystems (reviewed in (Bains *et al.* 2019a; Bains *et al.* 2019b; Glindemann *et al.* 1998; Roels *et al.* 2005; Roels and Verstraete 2001; Roels and Verstraete 2004):

1) $PH_3$ is directly produced by anaerobic bacteria from environmental phosphorus.
2) $PH_3$ is indirectly produced by anaerobic bacteria. Anoxic fermentation of organic matter by anaerobic bacteria results in acid products; these acid products, in turn, could react with inorganic metal phosphides, e.g. those present as trace elements in scrap metal, resulting in phosphine generation.

Of the two proposed explanations for biologically-associated production of phosphine, we argue that the direct production as a result of metabolic activity of anaerobic bacteria is the most likely. Our reasoning is based on two lines of evidence:

a) $PH_3$ has been detected in significant amounts in bacterial cultures in controlled laboratory experiments, where no metal phosphides were present, making the indirect acid-dependent production of $PH_3$ an unlikely scenario (Devai *et al.* 1988; Ding *et al.* 2005a; Ding *et al.* 2005b; Glindemann *et al.* 1996b; Jenkins *et al.* 2000; Liu *et al.* 2008; Schink and Friedrich 2000).

b) Several independent studies found that $PH_3$ was detected in feces from evolutionarily distant animals, inhabiting diverse environments, e.g. insects, birds and mammals (including humans) (e.g. (Chughtai and Pridham 1998; Gassmann and Glindemann 1993; Zhu *et al.* 2014)). It is implausible that there is a significant presence of contaminant metal phosphides in the guts of all the animals, which would be required for an indirect acid-dependent production of $PH_3$.

We end this introduction to the biological association of phosphine by noting that thermochemical studies on the feasibility of the production of $PH_3$ in temperate environments have found no plausible thermodynamically favored abiotic pathways, and as such $PH_3$ has no substantial false positives for life (see (Bains *et al.* 2019a), Section 4.3 and Appendix C). Conversely, production of $PH_3$ under anoxic conditions by living systems can be thermochemically favorable (Bains *et al.* 2019a) and biological functions that are accomplished through energy consuming reactions are not uncommon (Bains *et al.* 2019b); $PH_3$ could be used by life to perform complex functions that would warrant an energetic investment, such as signaling or a defense mechanism (Bains *et al.* 2019b). For more information about phosphine in the context of terrestrial biology and the thermodynamic feasibility of $PH_3$ production by life, see (Bains *et al.* 2019a; Bains *et al.* 2019b) and Appendices C, D and E.

## 2.3 Phosphine Chemistry in the Atmosphere

Within an atmosphere, phosphine is destroyed by the radicals O, H, and OH in reactions which are thought to be first-order with respect to its reactants and second-order overall. $PH_3$ can also be regenerated by reaction of $PH_2$ with H, and directly photolyzed by UV radiation. These processes are summarized below, and discussed in Section 5.2.

Reaction rate constants are expressed via the Arrhenius equation:

$$k = Ae^{-E/RT}$$

where $k$ is the reaction rate constant in units of $cm^3\ s^{-1}$, $A$ is a constant in units of $cm^3\ s^{-1}$, $E$ is the activation energy in units $J\ mol^{-1}$, $R$ is the gas constant in units of $J\ mol^{-1}\ K^{-1}$, and $T$ is temperature in K.

***Destruction by OH radicals*:** Oxidation with OH radicals is thought to be the main sink for phosphine in Earth's atmosphere via the reaction (Cao *et al.* 2000; Elm *et al.* 2017; Glindemann *et al.* 2005):

$$PH_3 + OH \rightarrow H_2O + PH_2$$
$$PH_2 + [O\ species] \rightarrow\rightarrow products\ (including\ H_3PO_4)$$

For this reaction, $A = 2.71 \times 10^{-11}\ cm^3\ s^{-1}$, $E = 1.29\ kJ\ mol^{-1}$, corresponding to $k_{PH3,OH} = 2 \times 10^{-11}\ cm^3\ s^{-1}$ at $T = 288$ K (Fritz *et al.* 1982).The lifetime of $PH_3$ due to OH reactions is calculated to be 28 hours at night and 5 hours in daytime, with the difference controlled by the concentration of UV-generated OH (Glindemann *et al.* 2003). The destruction of $PH_3$ by OH in the atmosphere eventually leads to phosphoric acid, which in turn contributes to the soluble phosphates found in rain water (Elm *et al.* 2017; Lewis *et al.* 1985).

***Destruction by O radicals:*** Phosphine also reacts very rapidly with atomic oxygen (on Earth, generated by photolyzed ozone), with reaction:

$$PH_3 + O \rightarrow\rightarrow products\ (including\ H_3PO_4).$$

For this reaction, $A = 4.75 \times 10^{-11}\ cm^3\ s^{-1}$, $E \approx 0$, corresponding to $k_{PH3,O} = 5 \times 10^{-11}\ cm^3\ s^{-1}$ at $T = 288$ K (i.e., temperature-independent in 208 - 423 K, Nava and Stief 1989). Because atomic oxygen is less abundant than OH in the Earth's atmosphere and troposphere, $PH_3$ destruction by OH is still the dominant route despite $PH_3$ reacting with O at a higher rate. On Earth, OH radicals described above are partially generated from the interaction between O radicals and water vapor, and so the reactions that produce the two radical species are not happening in isolation (Jacob 1999). In anoxic atmospheres, however, the main source of OH and H radicals is the photolysis of water vapor (Hu *et al.* 2012).

***Destruction by H radicals:*** Phosphine can be destroyed by the H radical via the reaction:

$$PH_3 + H \rightarrow H_2 + PH_2$$

For this reaction, $A$ = 7.22 x $10^{-11}$ $cm^3$ $s^{-1}$, $E$ = 7.37 kJ $mol^{-1}$, corresponding to $k_{PH_3,H}$=3 x $10^{-12}$ $cm^3$ $s^{-1}$ at $T$ = 288 K (Arthur and Cooper 1997). The reaction of $PH_3$ with the H radical is most relevant to $H_2$-rich atmospheres (Seager *et al.* 2013b).

***Recombination from H radicals:*** Phosphine can be regenerated by the radical recombination reaction:

$$PH_2 + H \rightarrow PH_3$$

With rate constant $k_{PH_2,H}$=3.7 x $10^{-10}$ exp(-340$K/T$) $cm^3$ $s^{-1}$, corresponding to $k_{PH_2,H}$=1.1 x $10^{-10}$ $cm^3$ $s^{-1}$ at $T$ = 288 K (Kaye and Strobel 1984). If [$PH_2$] is high, this reaction can be a major reformation pathway for $PH_3$ (see Section 5.2).

***Destruction through UV radiation***: UV radiation is thought to directly photolyze phosphine with unit quantum efficiency upon absorption of irradiation at wavelengths ≤ 230 nm (Kaye and Strobel 1984; Visconti 1981):

$$PH_3 + h\nu \rightarrow PH_2 + H$$

This photolysis reaction is not relevant on UV-shielded planets (e.g., modern Earth with its ozone layer), but could be relevant on anoxic planets where UV radiation may penetrate deeper into lower altitudes of the atmosphere.

Overall, phosphine is destroyed by UV irradiation, through both direct photolysis and through reactions with UV-generated radical species. However, $PH_3$ has been detected at concentrations of up to 7 ppt (2.45 ng $m^{-3}$) during daylight in Earth's high troposphere (Glindemann *et al.* 2003). $PH_3$ accumulates in the dry upper troposphere on Earth, because ozone attenuation of UV and lack of OH-producing $H_2O$ results in low abundances of OH radicals, which slows the $PH_3$ destruction and its return to the surface in the form of phosphates (Frank and Rippen 1987; Glindemann *et al.* 2003). The daytime-nighttime $PH_3$ concentration difference on Earth is large due to the generation of radicals by UV irradiation during the day, and their comparative absence at night.

***Solubility and aerosol formation***: Phosphine does not easily stick to aerosols and has very low water solubility (Fluck 1973). $PH_3$ is therefore a very mobile gas that is less likely to wash out and fall to the surface than other gases, such as hydrogen sulfide, methanethiol and ammonia (Glindemann *et al.* 2003).

UV photolysis of phosphine in the presence of hydrocarbons could lead to the formation of complicated alkyl-phosphines (Guillemin *et al.* 1997; Guillemin *et al.* 1995). Atmospheres that are prone to high concentrations of hydrocarbon radicals, such as $H_2$-rich atmospheres, could therefore lead to the creation of organophosphine hazes. The plausibility of organophosphine haze formation is discussed further in Section 5.2.

# 3. Inputs and Methods for the Assessment of Detectability

We have assessed the spectral distinguishability, atmospheric survival, and observational detectability of phosphine in anoxic exoplanets. In this section we first describe the choice of molecular inputs used for our spectral analyses (Section 3.1). We then provide a brief outline of the photochemical method used to calculate the distribution of molecules throughout the atmosphere (Section 3.2). Finally, we outline the method and detectability criteria for the simulations of observational spectra (Section 3.3).

## 3.1 Molecular Inputs

Molecular spectra can be represented in various forms to best serve as input for spectral representations and atmospheric models. For the comparison of phosphine with other major components of atmospheres (see Section 4.2) we have used cross-sections calculated from the most complete spectra available. The $PH_3$ molecular cross-sections come from the recently variationally computed $PH_3$ line list (Sousa-Silva *et al.* 2015) and the total internal partition function calculated in (Sousa-Silva *et al.* 2014). For all temperatures under 800 K (which includes all temperate environments), it is a complete $PH_3$ line list containing over 16 billion transitions between 7.5 million energy levels. Even at low temperatures, it is recommended that complete line lists are used for spectral simulations; complete line lists allow for more representative cross-sections with improved band shapes when compared to experimental or calculated spectra at room temperature. The carbon dioxide line list is from HITEMP (Rothman *et al.* 2010). All other molecular cross-sections are simulated using complete, theoretically calculated, line lists from the ExoMol database (Sousa-Silva *et al.* 2015; Tennyson *et al.* 2016; Yurchenko *et al.* 2011; Yurchenko and Tennyson 2014).

For the calculation of the transmission and thermal emission spectra of the model atmospheres (see Section 4.1) molecular opacities for phosphine are adopted from the ExoMol database (Sousa-Silva *et al.* 2015; Tennyson *et al.* 2016). For all other molecules we used the HITEMP and the HITRAN 2016 databases (Gordon *et al.* 2017; Rothman *et al.* 2010). Molecular cross-sections in the UV region, used to calculate photolysis rates (see Section 3.2), were obtained from the absorption cross-sections compendium of (Ranjan and Sasselov 2017) and from (Chen *et al.* 1991) via the MPI-Mainz Spectral Atlas (Keller-Rudek *et al.* 2013).

## 3.2. Photochemical Modelling

In this subsection we provide a brief description of the photochemical model used to calculate the concentration of phosphine as a function of altitude for a range of $PH_3$ surface fluxes. We also describe the atmospheric and stellar scenarios considered in our photochemical model. We found it necessary to use a photochemical model instead of the approximation of fixed radical profiles (e.g., Seager *et al.* 2013b) because of the intense reactivity of $PH_3$, which can drastically alter the radical profiles of an atmosphere. In particular, at high $PH_3$ fluxes the radical concentrations are suppressed due to reactions with $PH_3$, meaning that $PH_3$ can build up to much higher concentrations than a fixed radical profile approximation would predict.

### 3.2.1 Photochemical Model

We use the photochemical model of (Hu *et al.* 2012) to calculate atmospheric composition for different planetary scenarios. The model is detailed in (Hu et al., 2012); in brief, the code calculates the steady-state chemical composition of an exoplanetary atmosphere by solving the one-dimensional chemical transport equation. The model treats up to 800 chemical reactions, photochemical processes (i.e., UV photolysis of molecules), dry and wet deposition, surface emission, thermal escape of H and $H_2$, and formation and deposition of elemental sulfur and sulfuric acid aerosols. The model is designed to have the flexibility of simulating both oxidized and reduced conditions. Ultraviolet and visible radiation in the atmosphere is computed by the delta-Eddington two-stream method. The code has been validated by reproducing the atmospheres of modern Earth and Mars. The code and extensive application examples are described in several papers (Hu and Seager 2014; Hu *et al.* 2012; Hu *et al.* 2013; Seager *et al.* 2013a; Seager *et al.* 2013b). In calculating convergence, we required that the chemical variation timescale of significant species (>1 $cm^{-3}$) to be at least $10^{19}$ s, i.e. longer than the age of the universe.

We added phosphine to the model of (Hu *et al.* 2012). We considered surficial production as the only source of $PH_3$, and rainout, photolysis, and reactions with the main radical species O, H, and OH as the sinks (see Section 2.3). We take the Henry's Law constant for $PH_3$ from (Fu *et al.* 2013) via (Sander 2015). For photolysis, we take the $PH_3$ UV cross-sections at 295 K from (Chen *et al.* 1991). We follow (Kaye and Strobel 1984) in taking the branching ratio of this reaction to be unity, and take the quantum yield of $PH_3$ photolysis to be $q_\lambda = 1$ for $\lambda < 230$ nm and $q_\lambda = 0$ for $\lambda > 230$ nm. For the reactions with OH, O, and H, we take the rate constants from (Fritz *et al.* 1982), (Nava and Stief 1989), and (Arthur and Cooper 1997), as detailed in Section 2.3. We are unaware of geochemical constraints on the dry deposition velocity of $PH_3$; we take this value to be 0 cm $s^{-1}$, which could lead to an overestimation of $PH_3$ accumulation rates (see Section 5.2 for a discussion of possible $PH_3$ deposition). On the other hand, our approach neglects the possibility that atmospheric photochemistry may generate $PH_3$. In particular, we do not consider the recombination reaction $PH_2 + H \rightarrow PH_3$. This may lead to underestimating $PH_3$ accumulation, especially in $H_2$-dominated atmospheres where H abundances are high.

### 3.2.2 Planetary Scenarios

We model the atmospheres of Earth-sized, Earth-mass planets with two bulk atmospheric compositions: an $H_2$-dominated atmosphere and a $CO_2$-dominated atmosphere. We focus on $H_2$-dominated atmospheres because their low mean molecular masses make them amenable to characterization via transmission spectroscopy (Batalha *et al.* 2015). We focus on $CO_2$- dominated atmospheres as an oxidizing endmember to complement the reducing $H_2$-dominated case, and because early Earth is thought to have had a $CO_2$-rich atmosphere (Kasting 1993). We only consider anoxic atmospheres because $O_2$-rich atmospheres are likely to have large quantities of OH radicals which rapidly destroy $PH_3$ (see Section 2.3). Additionally, the aerobic metabolism of $O_2$-dependent life is likely to be incompatible with widespread $PH_3$ biological production (see Bains *et al.* 2019b).

Our atmospheres correspond to the $H_2$ and $CO_2$-dominated benchmark scenarios of (Hu *et al.* 2012), with the key difference that we do not set the rainout rates of $H_2$, CO, $CH_4$, $C_2H_6$, or $O_2$ to zero, as (Hu *et al.* 2012) did to simulate an abiotic planet. In brief, we consider planets with surface pressures of 1 bar, surface temperatures of 288K, and bulk dry atmospheric composition of 10% $N_2$, 90% $H_2$/$CO_2$ for the $H_2$/$CO_2$-dominated cases, respectively. The temperature profile is taken to evolve as a dry adiabat until 160K and 175K for the $H_2$ and $CO_2$-dominated cases, respectively, and isothermally thereafter. The strength of vertical mixing is scaled from that measured in Earth's atmosphere according to the mean molecular mass. The $H_2O$ concentration at the bottom of the atmosphere is set to 0.01, corresponding to 60% humidity. $H_2$, $CO_2$, $CH_4$, $SO_2$, and $H_2S$ are emitted from the surface at rates corresponding to terrestrial volcanism. The mixing ratio profile of the dominant gases (gases with abundances exceeding 100 ppb) used for the modeling of the $H_2$-rich atmosphere on a massive super Earth orbiting an active M-dwarf is shown in Figure 2 (see Appendix A for the mixing ratio profiles used to model the remaining atmospheric scenarios). For further details, including the rationale for these parameters, see (Hu *et al.* 2012).

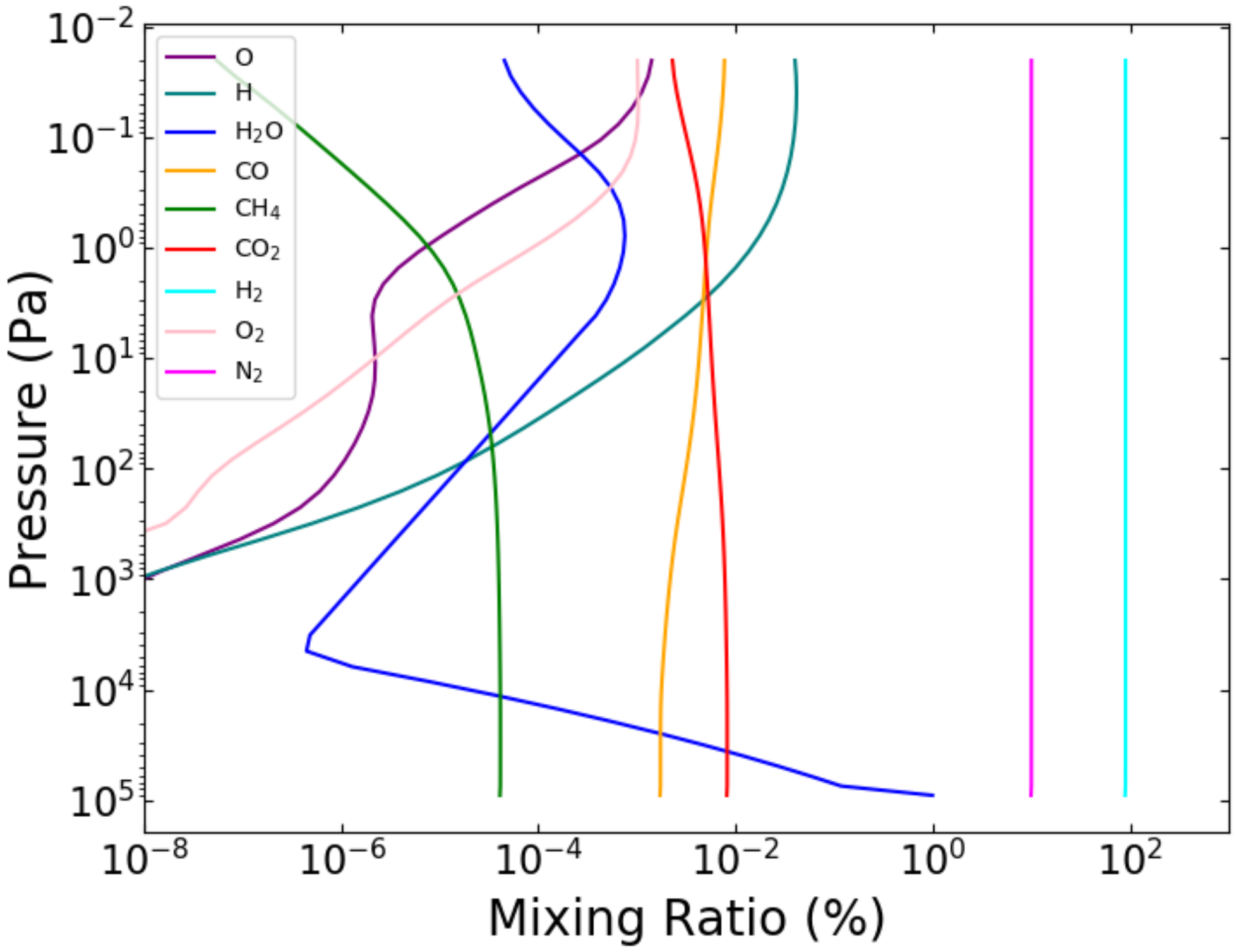

**Figure 2.** Mixing ratio profile of a $H_2$-rich atmosphere on an Earth-sized planet orbiting an active M-dwarf. Vertical axis represents pressure in units of Pa and the horizontal axis shows the mixing ratio represented as a percentage of the total atmospheric layer. Figure partially adapted from (Hu *et al.* 2012) and (Seager *et al.* 2013b).

Stellar irradiation is a key input for photochemical models. We considered instellation corresponding to the Sun (our "Sun-like" case; (Hu *et al.* 2012)) and from the M-dwarf GJ1214 (our "active M-dwarf" case; (Seager *et al.* 2013b)). The semi-major axes of the planets for the ($H_2$, $CO_2$)-dominated case are taken to be (1.6 A, 1.3 AU) for the Sun-like case, and (0.042 AU, 0.034 AU) for the "active M-dwarf" case, corresponding to surface temperatures of 288K at 0 $PH_3$ emission.

For a sensitivity test we also considered a theoretical "quiet" M-dwarf simulated by reducing the UV-flux of GJ1214 by three orders of magnitude (corresponding to approximately a factor of 100 less UV radiation than the least active M-dwarf known GJ581 (France *et al.* 2016). See section 4.1.4 for a discussion of the sensitivity of our results to changes in surface temperature and to low UV irradiation levels.

## 3.3. Atmospheric Spectral Simulations

We use the outputs of the photochemical models described above to model observational spectra projected to massive super-Earths with $M_p$ =10 $M_E$ and $R_p$ = 1.75 $R_E$. We focus on such large, massive planets due to the following observational considerations: 1) such planets are easier to detect via radial velocities and transit observations; 2) large planets have larger thermal emission signatures; and 3) massive planets are more likely to retain $H_2$-rich atmospheres, which are much easier to characterize in transmission because of their larger scale heights than other atmosphere types.

### 3.3.1 Transmission and Emission Spectral Calculations

Transmission and thermal emission spectra were simulated with the program SEAS (Simulated Exoplanet Atmosphere Spectra). The projection from Earth-sized to super-Earths is performed using equivalent techniques to those in (Hu *et al.* 2012); SEAS takes as input a list of molecular mixing ratios as a function of pressure, which are invariant to first-order to changes in the surface gravity.

The transmission spectrum code calculates the optical depth along the limb path, and sums up chords assuming the planet atmosphere is homogenous. The SEAS transmission spectrum code is similar in structure to that described in both (Kempton *et al.* 2017; Miller-Ricci *et al.* 2009), with the main difference being that SEAS can accept variable mixing ratio inputs, important for super-Earths whose atmospheres are severely impacted by photochemistry. The temperature-pressure profile, including limits and resolution, is specified by the user. Molecular line lists are taken from HITRAN 2016 and ExoMol (Gordon *et al.* 2017; Tennyson *et al.* 2016), with cross sections calculated with HAPI (Kochanov *et al.* 2016) and ExoCross (Yurchenko *et al.* 2018), respectively. The molecular species are chosen by the user, and all molecules in the HITRAN and ExoMol databases are user-selectable options.

The thermal emission code integrates a blackbody exponentially attenuated by the optical depth without scattering (e.g., (Seager 2010)). The code uses the same input temperature-pressure profile and molecular cross-sections as described above. SEAS considers clouds in the emergent spectra for thermal emission by averaging cloudy and cloud-free spectra (resulting in 50% cloud coverage). We omit clouds or hazes for the transmission spectra model; if the atmosphere is cloudy or hazy at high altitudes, the spectral features in transmission will be muted. Consequently, our calculations represent upper bounds on the magnitude of the transmission features with respect to cloud or haze effects. We discuss the impact of clouds in our modelled transmission spectra in Section 5.2.

The SEAS transmission code has been validated by comparing results to the Atmospheric Chemistry Experiment data set (Bernath *et al.* 2005) for transmission spectrum and the MODTRAN spectrum (Berk *et al.* 1998) for thermal emission spectrum. We also compared results related to this phosphine work to transmission spectra generated by the code described in (Hu *et al.* 2012).

### 3.3.2 Detectability Metric

We study the spectroscopic detectability of phosphine in $H_2$- and $CO_2$-rich atmospheres, in transmission and emission observation scenarios. In transmission we compare to the mean transit depth of the planet radius, and in emission to the blackbody curve. We consider a 6.5 m space telescope, having a quantum efficiency of 25% observing with a 50% photon noise limit. We consider our 1.75 $R_{Earth}$-planet to be orbiting 1) a 0.26-$R_{sun}$ M dwarf star at 5 pc with an effective temperature of 3000 K; and 2) a Sun-like star. Stellar flux is the source of the noise, combining in-transit and out-of-transit flux noises. The theoretical transmission spectra are based on JWST and its NIRSpec and MIRI instruments (Bagnasco *et al.* 2007; Wright *et al.* 2010). To calculate theoretical thermal emission spectra, we consider a secondary eclipse scenario as observed from JWST with the MIRI instruments (both mid- and low-resolution spectrometers). We binned the data to a resolution of R ~ 10 to increase the significance of detection.

We investigate the detectability of phosphine in exoplanet atmospheres by adapting the detection metrics defined in (Seager *et al.* 2013b) and (Tessenyi *et al.* 2013). The detectability metric is a theoretical metric using simulated data.

We first simulate model independent observational data for all planetary scenarios considered (e.g., using the instrumental constraints of JWST). For an analysis of the transmission spectra models, we then compare the wavelength-dependent transit depth of the planet to the “white-light” transit depth in each wave-band (corresponding to the coverage of each instrument). Phosphine is considered detectable if we can detect opacity at wavelengths corresponding to $PH_3$ absorption features with statistically significant confidence. To establish the statistical significance of opacity detection in transmission, we assume a simulated spectrum and then assign binned values for the transit depth. We then calculate the wavelength-dependent one-sigma (1-$\sigma$) error bar for each binned value (i.e. standard deviation) using only stellar photon noise. The significance of the deviation is calculated with:

$$\frac{|\mu_\lambda - \mu_{\bar{\lambda}}|}{\sqrt{\sigma_\lambda^2 + \sigma_{\bar{\lambda}}^2}},$$

where $\mu_\lambda$ is the wavelength-dependent transit depth of the simulated atmosphere, $\mu_{\bar{\lambda}}$ is the mean transit depth of the white-light averaged waveband, and σ is the uncertainty on the measurement. The uncertainties are estimated based on shot noise. We then assess the detectability of phosphine by simulated a model atmosphere with and without $PH_3$, and comparing the deviation of each atmosphere from their associated mean. This comparison establishes whether a model atmosphere with $PH_3$ fits the simulated observational data better than one without $PH_3$.

In thermal emission, we use a similar detectability metric to the transmission analysis described above, with the distinction that we calculate the deviation of our modelled atmosphere spectra from its best-fit blackbody continuum (instead of the white light average used for transmission comparisons). The temperature of the blackbody is set by fitting a blackbody curve to the simulated data.

The integration time is a variable parameter in the SEAS models, but features are only considered detectable if they achieve at least a 3-$\sigma$ interval with 200 observation hours or less (considering 100 hours in-transit and 100 hours out-of-transit).

### 3.3.3 Scaling to Smaller Planets

We performed our spectral simulations for a massive super-Earth planet ($M_p$ =10 $M_E$ and $R_p$ = 1.75 $R_E$). In this section we consider how the prospects for atmospheric characterization scale to smaller, more Earth-sized worlds.

The amplitude of the atmospheric absorption signal in transmission is characterized by (Brown 2001):

$$\frac{\delta A}{A} = \frac{2\pi R_p(kT/\mu g)}{\pi R_*^2} = \frac{2\pi R_p(kTR_p^2/\mu G M_p)}{\pi R_*^2} \propto \frac{R_p^3}{M_p},$$

where $R_p$ is the planet radius and $M_p$ is the planet mass. This implies that the transmission spectroscopy signal from a 1 $R_E$, 1 $M_E$ planet should be twice the signal from the 1.75 $R_E$, 10 $M_E$ planet we consider here, and the $PH_3$ surface fluxes required to produce a detectable atmospheric signal should be half of what we model for our super-Earth scenario[4].

The amplitude of the thermal emission spectral signal is characterized by:

$$F(\lambda) \propto [B(T_{cont}, \lambda) - B(T_{line}, \lambda)]R_p^2,$$

[4] We note that it is uncertain whether a small planet can retain a $H_2$-dominated atmosphere over geological time in the face of atmospheric escape.

where $\lambda$ is the wavelength, $B$ is the blackbody function, and $T_{cont}$ and $T_{atm}$ are, respectively, the brightness temperature in and out of the spectral line under consideration. The above equation implies that the thermal emission signal from a 1 $R_E$, 1 $M_E$ planet should be a third of the signal from the 1.75 $R_E$, 10 $M_E$ planet we consider here, and the $PH_3$ surface fluxes required to produce a detectable atmospheric signal should be three times larger than what we model for our super-Earth scenario.

We conclude that spectrally characterizing Earth-sized planets is comparable in difficulty to characterizing super Earth planets, but that characterizing the atmospheres of smaller worlds is somewhat easier in transmission and somewhat harder in emission. The differences do not affect our main conclusions.

# 4. Results

We find that phosphine as a detectable biosignature gas has three encouraging properties: (1) $PH_3$ can accumulate to detectable levels in an exoplanet atmosphere, provided it has a high production rate at the planet's surface (Section 4.1); (2) $PH_3$ has unique spectral features, namely strong bands around 2.7-3.6 microns, 4.0-4.8 microns and 7.8-11.5 microns, that allow it to be distinguishable from other dominant atmosphere molecules (Section 4.2); and (3) based on the abundances and surface fluxes needed to produce detectable levels of $PH_3$, it has no known false positives provided that the planet's surface temperature is below 800 K (Section 4.3). In addition to the above our results show that, at surface fluxes near the minimum flux necessary to allow for $PH_3$ detection, a "runaway" effect for $PH_3$ occurs (Section 5.1)

We present results for the phosphine detection in $H_2$-rich and $CO_2$-rich atmospheres, for planets orbiting Sun-like stars and active M-dwarf stars.

## 4.1. Phosphine Detection in Exoplanet Atmospheres

We performed a series of simulations and calculations to explore the prospects for detecting phosphine in an exoplanet atmosphere using transmission and thermal emission spectroscopy. We consider $H_2$ and $CO_2$-dominated atmospheres, and stellar irradiation environments corresponding to the modern Sun ("Sun-like") and GJ1214 ("Active M-dwarf"); see (Seager *et al.* 2013b) for details.

In this section, we first present a set of simulated spectra, both in transmission (4.1.1) and emission (4.1.2), for atmospheric scenarios with and without phosphine. Here we predict the minimum abundances required for $PH_3$ to be detectable in each atmosphere considered. We calculate the required surface production rates for $PH_3$ to achieve the abundances required for detection in Section 4.1.3, using our photochemical model to simulate the equilibrium distribution of atmospheric gases throughout each atmosphere. Finally, we assess the sensitivity of our results to a variable surface temperature and a host star with low levels of radiation (Section 4.1.4).

### 4.1.1 Amount of Phosphine Required for Detection via Transmission Spectroscopy

We find that $PH_3$ is detectable in anoxic atmospheres only if it is able to accumulate to the order of ppb to 100s of ppm, for $H_2$-rich and $CO_2$-rich atmospheres, respectively. For comparison, $PH_3$ is present at the ppt to ppb level on modern Earth. We estimate the photochemical plausibility of $PH_3$ accumulating to such large abundances in Section 4.1.3.

Unfortunately, even with high concentrations of phosphine in the atmosphere, many tens of hours of JWST time are needed. The atmosphere mixing ratios, the surface production rates required, and the number of observation hours needed for different planet and host star scenarios are provided in Table 1.

| **Atmospheric Scenario** | **Required Mixing Ratio for Detection** | **Minimum Observation Hours (in-transit + out-of-transit)** | **Associated Confidence Interval for Phosphine Detection (σ)** |
|---|---|---|---|
| $H_2$-dominated, Sun-like star | 780 ppm | 56 | 3 |
| $H_2$-dominated, active M-dwarf (Fig. 3) | 220 ppb | 91 | 3 |
| $H_2$-dominated, active M-dwarf | 220 ppb | 200 | 4.4 |
| $H_2$-dominated, active M-dwarf | 5 ppb | 200 | 2.5 |
| $H_2$-dominated, active M-dwarf | *0.28%* | *3* | *5* |
| $CO_2$-dominated, Sun-like star | N/A | Not detectable | N/A |
| $CO_2$-dominated, active M-dwarf (Fig. 4) | 310 ppm | 200 | 2.7 |
| $CO_2$-dominated, active M-dwarf | *7.6%* | *32* | *3* |

**Table 1**: Phosphine mixing ratios needed for detection in transmission for different atmospheric and stellar scenarios, with associated observation and surface flux requirements. For planets orbiting an active M-dwarf, $PH_3$ requires minimum abundances of 220 ppb and >310 ppm to be detectable with a 3-σ confidence interval on $H_2$- and $CO_2$-rich atmospheres, respectively. Values in italic correspond to atmospheric scenarios where $PH_3$ is at a runaway threshold (see Figure 5, and Sections 4.1.3 and 5.1). For planets orbiting a Sun-like star, $PH_3$ must become a major component of the atmosphere for its detection to be possible with less than 200 observation hours or, in the case of $H_2$-rich atmospheres, with surface production rates above those found anywhere on Earth (i.e. >$10^{14}$ $cm^{-2}$ $s^{-1}$). For comparison, biological production of $CH_4$ on Earth corresponds to $1.2 \times 10^{11}$ $cm^{-2}$ $s^{-1}$, of which a significant proportion is due to human activity (Guzmán-Marmolejo and Segura 2015; Houghton 1995; Segura *et al.* 2005).

Planets with $H_2$-rich atmospheres orbiting active M-dwarfs require the smallest phosphine abundances for its detection (10s to 100s of ppb; see Figure 3), which can be expected due to their lower mean radical concentrations compared to an oxidized atmosphere (Seager *et al.* 2013b). $H_2$-rich atmospheres also have transmission spectra that are easier to detect than planets with higher mean molecular weight atmospheres (e.g., $CO_2$) because of their larger scale height, i.e. a “puffier” atmosphere.

For $H_2$-rich atmospheres, only the strongest band of phosphine at 4.0-4.8 microns can be detectable (Figure 3). The other $PH_3$ features are either too weak or contaminated by other atmospheric molecular species.

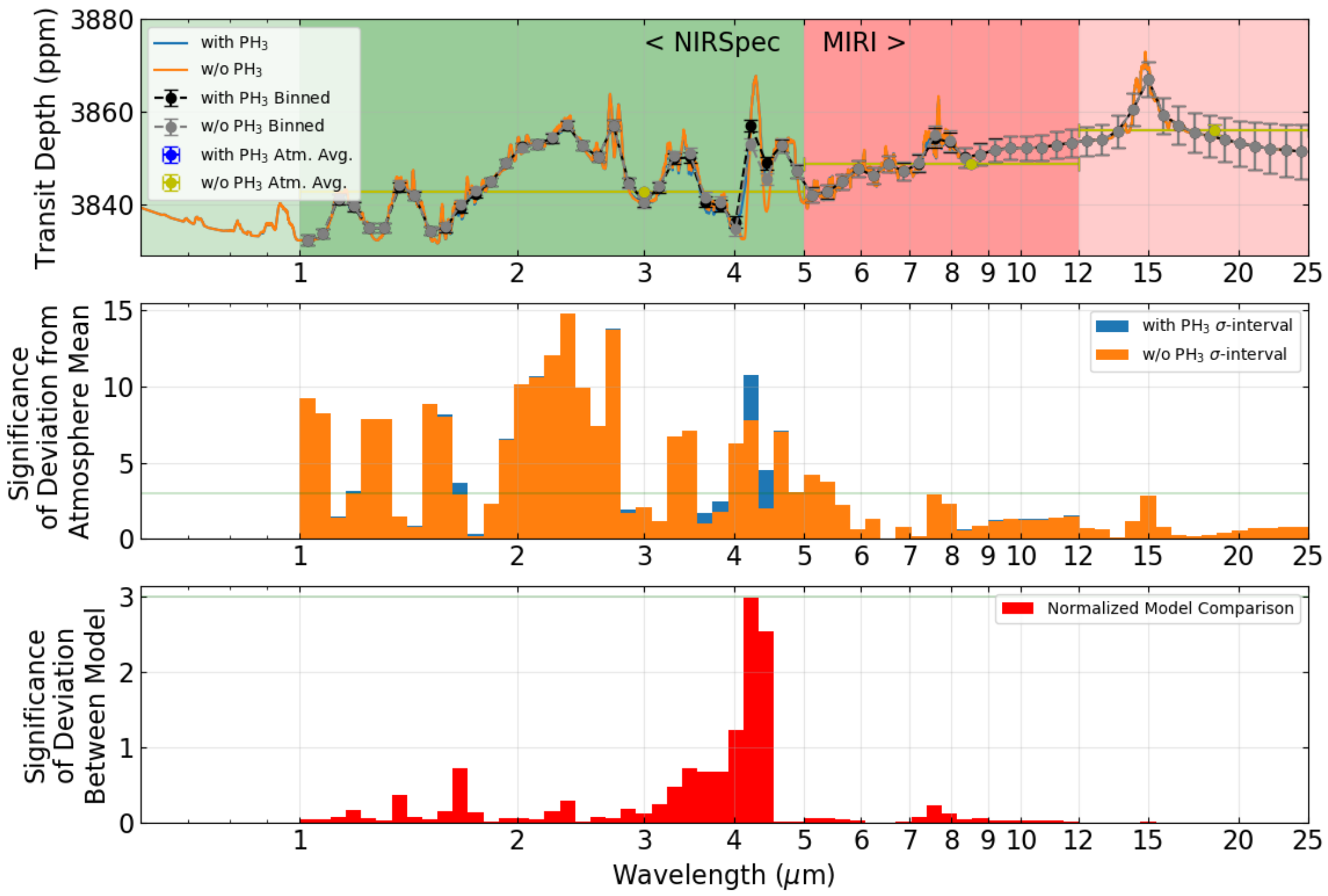

**Figure 3:** Theoretical transmission spectra for an $H_2$-rich atmosphere on a 10 $M_E$, 1.75 $R_E$ planet with a surface temperature of 288 K orbiting an active M-dwarf (1 bar atmosphere composed of 90% $H_2$ and 10% $N_2$), after 91 hours of observation. **Top panel**: Vertical axis shows transit depth of the simulated atmosphere spectra in units of ppm (y-axis) and horizontal axis showing wavelength in microns. The orange curve corresponds to the simulated atmosphere spectrum without $PH_3$, and the blue curve to an atmosphere spectrum with $PH_3$, simulated considering a $PH_3$ concentration of 220 ppb. Blue error bars correspond to the wavelength-averaged uncertainty within the instrumental waveband; black and gray error bars correspond to the uncertainty of each wavelength bin for atmosphere models with and without $PH_3$, respectively. Green and pink shading represent the wavelength coverage of the NIRSpec and MIRI instruments (Bagnasco *et al.* 2007; Wright *et al.* 2010). **Middle panel:** Vertical axis shows the statistical significance of detection for two different model atmospheres, with $PH_3$ (blue) and without $PH_3$ (orange). **Bottom panel**: Statistical significance of the detection of $PH_3$ opacities at each wavelength bin. Vertical axis shows size of the statistical deviation between atmosphere models with and without $PH_3$ (units of σ-interval). In the middle and bottom panels the horizontal green line represents the 3-σ statistical significance threshold, and the horizontal axes show the individual wavelength bins (microns). For $H_2$-dominated atmospheres the 4.0 - 4.8 microns spectral band of $PH_3$ is the most promising feature for detection in transmission.

For $CO_2$-rich atmospheres (Figure 4), several spectral bands of phosphine show substantial spectral absorptions when compared to atmospheres without $PH_3$. Nonetheless, no spectral band of $PH_3$ can achieve a 3-σ statistical significance even after 200 observation hours when considering a planet orbiting an active M-dwarf.

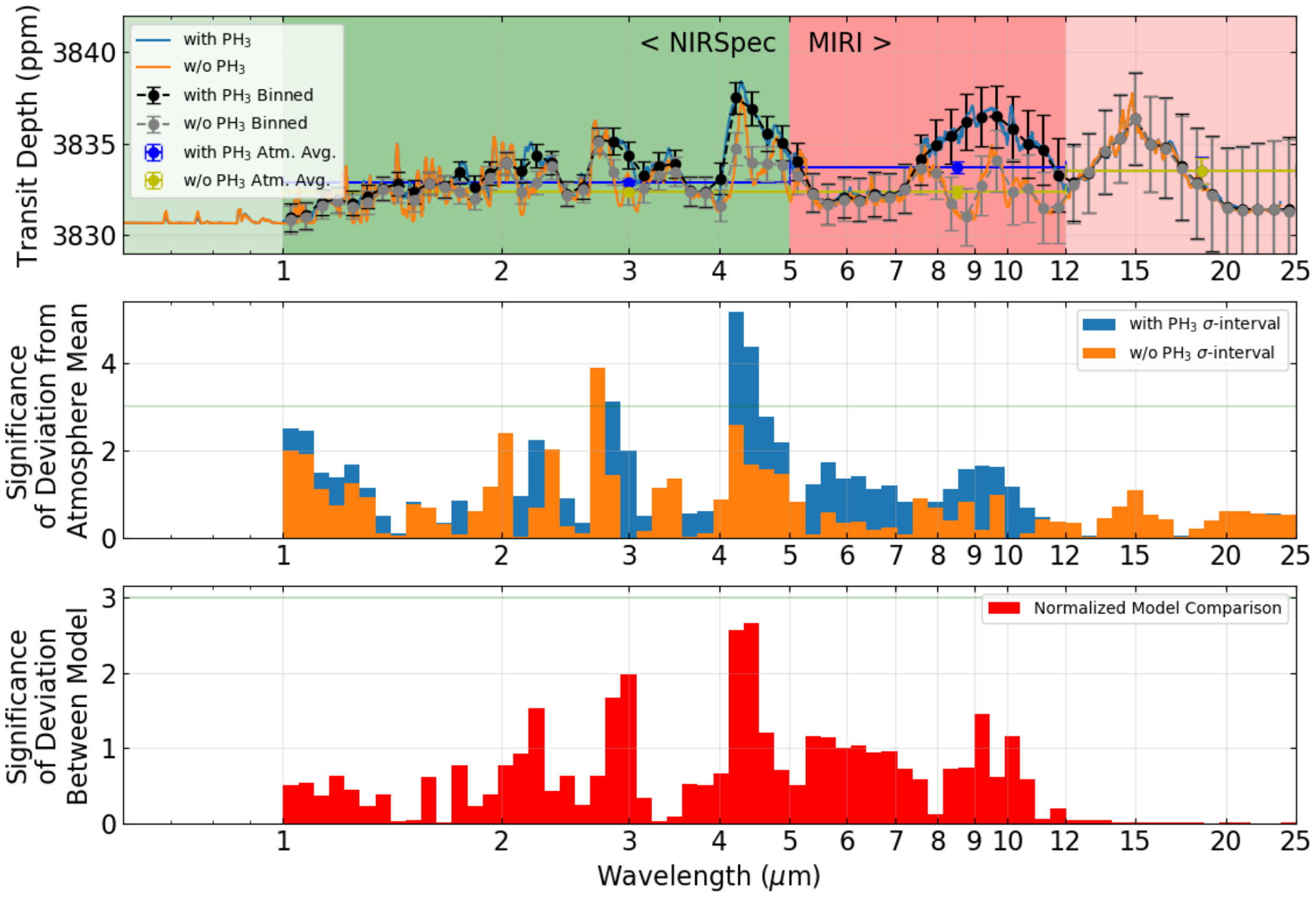

**Figure 4:** Theoretical transmission spectra for an $CO_2$-rich atmosphere on a 10 $M_E$, 1.75 $R_E$ planet with a surface temperature of 288 K orbiting an active M-dwarf (1 bar atmosphere composed of 90% $CO_2$ and 10% $N_2$), after 200 hours of observation. **Top panel**: Vertical axis shows transit depth of the simulated atmosphere spectra in units of ppm (y-axis) and horizontal axis showing wavelength in microns. The orange curve corresponds to the simulated atmosphere spectrum without $PH_3$, and the blue curve to an atmosphere spectrum with $PH_3$, simulated considering a $PH_3$ concentration of 310 ppm. Blue error bars correspond to the wavelength-averaged uncertainty within the instrumental waveband; black and gray error bars correspond to the uncertainty of each wavelength bin for atmosphere models with and without $PH_3$, respectively. Green and pink shading represent the wavelength coverage of the NIRSpec and MIRI instruments (Bagnasco *et al.* 2007; Wright *et al.* 2010). **Middle panel:** Vertical axis shows the statistical significance of detection for two model atmospheres, with $PH_3$ (blue) and without $PH_3$ (orange). **Bottom panel**: Statistical significance of the detection of $PH_3$ opacities at each wavelength bin. Vertical axis shows size of the statistical deviation between atmosphere models with and without $PH_3$ (units of σ-interval). In the middle and bottom panels the horizontal green line represents the 3-σ statistical significance threshold, and the horizontal axes show the individual wavelength bins (microns). In $CO_2$-dominated atmospheres several spectral features of $PH_3$ have substantial opacities, but no feature achieves a 3-σ statistical significance when compared to the model atmosphere without $PH_3$.

Phosphine is very difficult to detect on planets orbiting Sun-like stars. Planets with $CO_2$-dominated atmospheres require longer than 200 observation hours for the detection of $PH_3$ in transmission spectra, even with the highest surface fluxes considered ($3x10^{13}$ $cm^{-2}$ $s^{-1}$). The detection of $PH_3$ can only achieve a 3-σ statistical significance on planets with $H_2$-rich atmospheres for fluxes of $10^{14}$ $cm^{-2}$ $s^{-1}$ (Table 1); this flux is comparable to the highest recorded $PH_3$ flux on Earth (above sewage plants; Devai *et al.* 1988) and above the values for the biological production of methane, which on Earth corresponds to $1.2x10^{11}$ $cm^{-2}$ $s^{-1}$ (Guzmán-Marmolejo and Segura 2015; Segura *et al.* 2005).

The results presented above show that it is possible, but difficult, to detect phosphine in anaerobic atmospheres if it is present as a trace gas. However, if $PH_3$ production rates increase sufficiently, they outpace the ability of stellar

NUV photons to destroy $PH_3$, whether via direct photolysis or via generation of radical species. $PH_3$ may then become a significant component of the atmosphere (e.g., hundreds to thousands of ppm), and its detectability increases dramatically. The $PH_3$ surface fluxes required to reach this runaway phase (~$10^{12}$ $cm^{-2}$ $s^{-1}$) are not significantly higher than those required for detection (~$10^{10-11}$ $cm^{-2}$ $s^{-1}$). For example, with surface production rates only 9 times larger than those that produce the atmospheric spectrum shown in Figure 3, $PH_3$ reaches the runaway threshold, and can be detected with 5-σ statistical significance after only 3 hours of observation (see Figure 5). The plausibility of this runaway effect is discussed further in Sections 4.1.3 and 5.1.

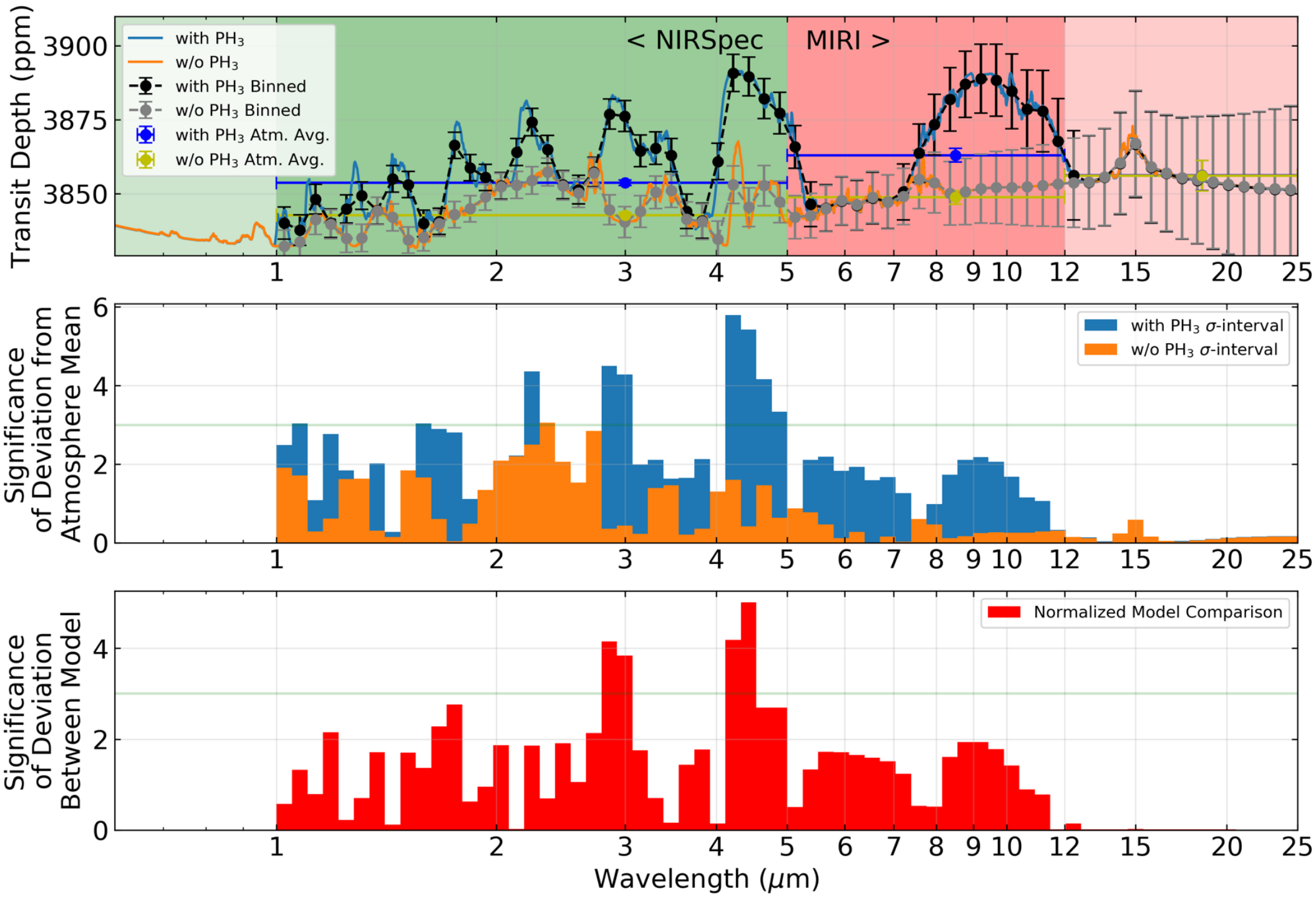

**Figure 5**: Theoretical transmission spectra for an $H_2$-rich atmosphere on a 10 $M_E$, 1.75 $R_E$ planet with a surface temperature of 288 K orbiting an active M-dwarf (1 bar atmosphere composed of 90% $H_2$ and 10% $N_2$), at the threshold of the phosphine runaway phase. **Top panel**: Vertical axis shows transit depth of the simulated atmosphere spectra in units of ppm (y-axis), after 3 hours of observation, and horizontal axis showing wavelength in microns. The orange curve corresponds to an atmosphere spectrum without $PH_3$, and the blue curve to an atmosphere spectrum with $PH_3$, simulated considering a $PH_3$ concentration of 0.28%. Blue error bars correspond to the wavelength-averaged uncertainty within the instrumental waveband; black and gray error bars correspond to the uncertainty of each wavelength bin for atmosphere models with and without $PH_3$, respectively. Green and pink shading represent the wavelength coverage of the NIRSpec and MIRI instruments (Bagnasco *et al.* 2007; Wright *et al.* 2010). **Middle panel:** Vertical axis shows the statistical significance of detection for two model atmospheres, with $PH_3$ (blue) and without $PH_3$ (orange). **Bottom panel**: Statistical significance of the detection of $PH_3$ opacities at each wavelength bin. Vertical axis shows size of the statistical deviation between atmosphere models with and without $PH_3$ (units of σ-interval). In the middle and bottom panels the horizontal green line represents the 3-σ statistical significance threshold, and the horizontal axes show the individual wavelength bins (microns). Once $PH_3$ enters the runaway phase it can be detected after a few hours of observations, through its two strong features in the 2.7 – 3.6 and 4 - 4.8 microns regions.

### 4.1.2 Amount of Phosphine Required for Detection via Emission Spectroscopy

We now examine the influence of phosphine in the simulated emission spectra of $H_2$- and $CO_2$-rich planets orbiting Sun-like stars and active M-dwarfs. Our findings for the amount of $PH_3$ required for detection in thermal emission are similar to that in transmission, i.e. $PH_3$ can only be detected with many tens of hours of observation time (Table 2). We find that, in emission, $PH_3$ is detectable in anoxic atmospheres only if it is able to accumulate to at least abundances in the order of ppb; for comparison, $PH_3$ is present at the ppt to ppb level on modern Earth. The photochemical plausibility of $PH_3$ accumulating to such large abundances is presented in Section 4.1.3.

| **Atmospheric Scenario** | **Required Mixing Ratio for Detection** | **Minimum Observation Hours (in-transit + out-of-transit)** | **Associated Confidence Interval for Phosphine Detection (σ)** |
|---|---|---|---|
| $H_2$-dominated, Sun-like star | N/A | Not detectable | N/A |
| $H_2$-dominated, active M-dwarf (Fig. 6) | 220 ppb | 131 | 3 |
| $H_2$-dominated, active M-dwarf | 4 ppm | 52 | 3 |
| $CO_2$-dominated, Sun-like star | N/A | Not detectable | N/A |
| $CO_2$-dominated, active M-dwarf (Fig. 7) | 15 ppm | 150 | 3 |
| $CO_2$-dominated, active M-dwarf | 310 ppm | 48 | 3 |

**Table 2**: Phosphine mixing ratios needed for detection in emission for different atmospheric and stellar scenarios, with associated observation and surface flux requirements. For planets orbiting an active M-dwarf, $PH_3$ requires minimum abundances of 220 ppb and 15 ppm to be detectable on $H_2$- and $CO_2$-rich atmospheres, respectively. For planets orbiting a Sun-like star, no scenario where $PH_3$ is not a major component of the atmosphere could allow for its detection with less that 200 observation hours.

Our models show that the detection of a $CO_2$- or $H_2$-rich atmosphere with high statistical significance (>5-σ) is feasible with only a few tens of observation hours. However, the unambiguous attribution of opacity to phosphine requires much longer observation times (see Table 2). As an observer, a detection can be considered as an offset to the blackbody curve but these are only reliable if the blackbody temperature has been accurately estimated. Our detection metric uses a blackbody curve created from a best fit to the simulated observations, which biases towards low-temperatures by non-$PH_3$ absorbers. In reality, the temperature of the planet may be obtained through other methods, so our results can be considered a conservative estimate for the minimum $PH_3$ abundances required for detection.

We find the most detectable spectral region of phosphine in thermal emission is the broad band at 7.8 – 11.5 microns (Figures 6 and 7). In emission, planets orbiting an active M-dwarf require the smallest $PH_3$ abundances (100s of ppb to 100s of ppm) to confirm its detection, achieving a 3-σ confidence interval with a minimum of 52 and 48 hours of observation, for $H_2$- and $CO_2$-rich atmospheres, respectively.

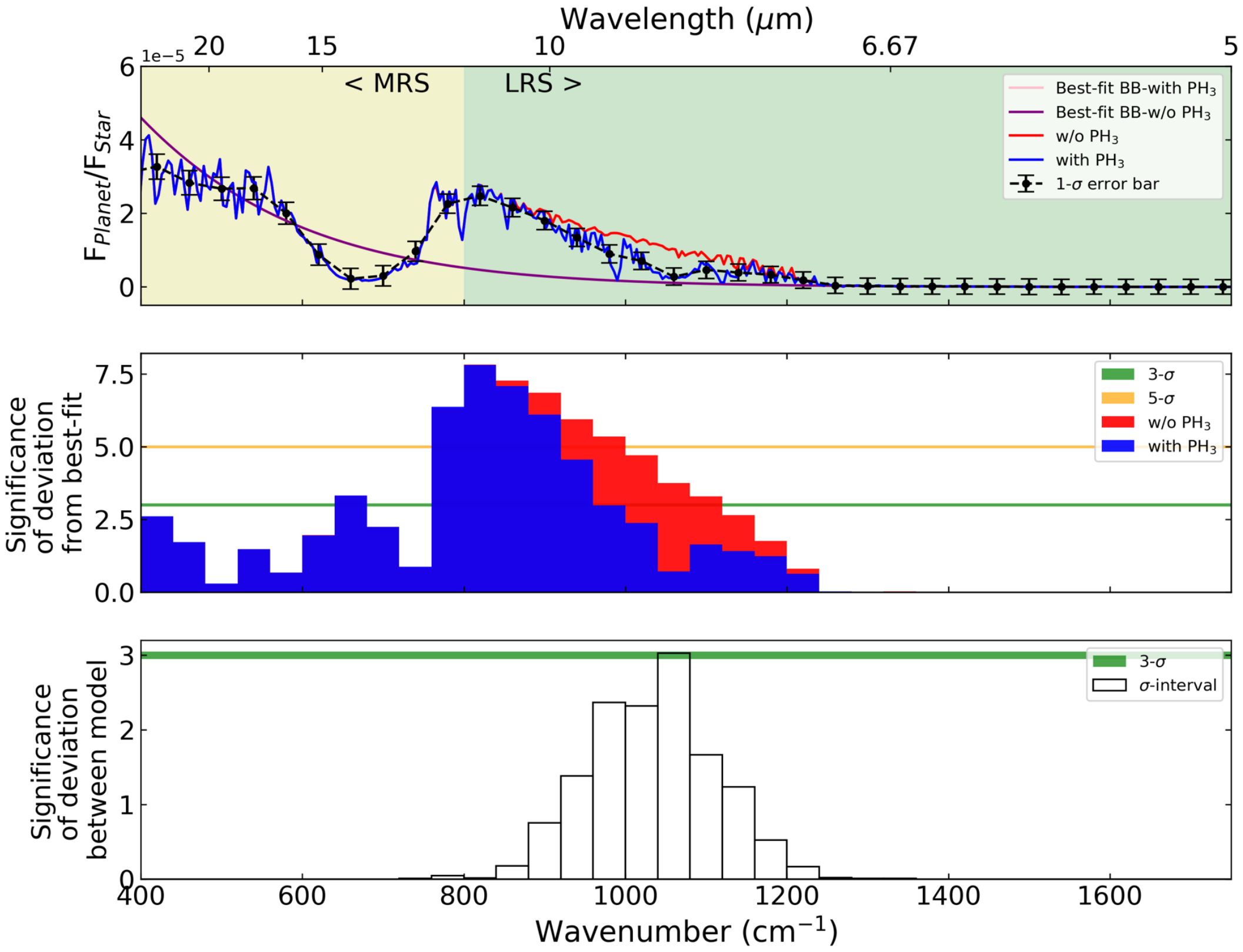

**Figure 6.** Detectability of phosphine in the emission spectrum of a super Earth exoplanet (10 $M_E$ and 1.75 $R_E$) with a $H_2$-rich atmosphere orbiting an active M-dwarf, after 131 hours of observation. Horizontal axes show wavelength in microns (top) and wavenumbers in inverse $cm^{-1}$ (bottom). **Top panel**: vertical axes show the flux ratio between the star and the planet; pink and purple lines represent the blackbody curves fitted to the simulated observational data for atmospheres with and without $PH_3$, respectively; blue and red curves represent a modelled atmosphere with a $PH_3$ mixing ratio of 220 ppb and an atmosphere without $PH_3$, respectively; black error bars represent the 1-σ uncertainty in the observed data; MRS (yellow shading) and LRS (green shading) represent the coverage of the JWST mid- and low-resolution MIRI instruments, respectively. **Middle panel**: Statistical significance of the detection of an atmosphere with (blue) and without (red) $PH_3$ when compared to their best-fit blackbody curves, in units of σ-interval; the horizontal green and orange lines represent the 3-σ and 5-σ statistical significance threshold, respectively. **Bottom panel**: Statistical significance of the deviation between an atmospheric model with and without $PH_3$; the horizontal green line represents the 3-σ statistical significance threshold. The detection of $PH_3$ achieves a 3-σ confidence interval, through the high frequency wing of its strong broad band at 7.8 – 11.5 microns.

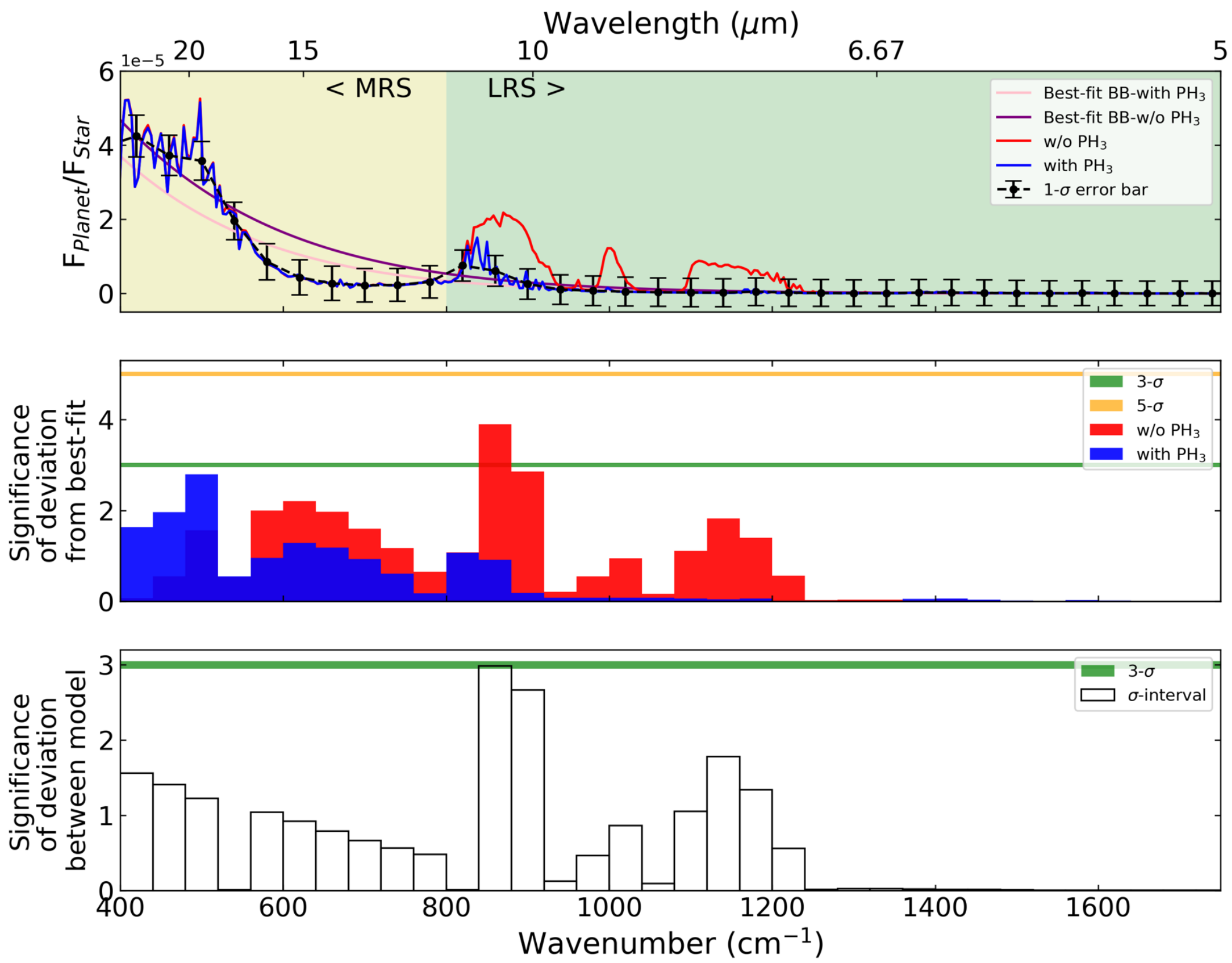

**Figure 7.** Detectability of phosphine in the emission spectrum of a super Earth exoplanet (10 $M_E$ and 1.75 $R_E$) with an $CO_2$-rich atmosphere orbiting an active M-dwarf, after 48 hours of observation. Horizontal axes show wavelength in microns (top) and wavenumbers in inverse $cm^{-1}$ (bottom). **Top panel**: vertical axes show the flux ratio between the star and the planet; pink and purple lines represent the blackbody curves fitted to the simulated observational data for an atmosphere with and without $PH_3$, respectively; blue and red curves represent a modelled atmosphere with a $PH_3$ mixing ratio of 310 ppm and an atmosphere without $PH_3$, respectively; black error bars represent the 1-sigma uncertainty in the observed data; MRS (yellow shading) and LRS (green shading) represent the coverage of the JWST mid- and low-resolution MIRI instruments, respectively. **Middle panel**: Statistical significance of the detection of an atmosphere with (blue) and without (red) $PH_3$ when compared to their best-fit blackbody curves, in units of σ-interval; the horizontal green and orange lines represent the 3-σ and 5-σ statistical significance threshold, respectively. **Bottom panel**: Statistical significance of the deviation between an atmospheric model with and without $PH_3$; the horizontal green line represents the 3-σ statistical significance threshold. The detection of $PH_3$ achieves a 3-σ confidence interval, through its strong broad band at 7.8 – 11.5 microns.

Detection of phosphine on planets orbiting Sun-like stars is difficult. In these scenarios the detection of any modelled super-Earth atmosphere cannot achieve a 3-σ confidence interval even with 200 observation hours.

We note that at sufficiently high phosphine concentrations, our model shows that the wings of the $PH_3$ absorption features become opaque (e.g., the strong, broad, band at 7.8 -11 microns) and our emission spectra probe the isothermal stratosphere. Consequently, if $PH_3$ concentrations are high enough, our models show that it is not possible to detect wavelength-dependent opacities due to $PH_3$ on the basis of emission data alone. At face value, this observation implies a maximum $PH_3$ concentration and flux past which it is impossible to detect $PH_3$ in emission. However, in reality this effect is an artefact of our assumption of an isothermal stratosphere. The stratosphere may have temperature variations, which would facilitate the detection of wavelength-dependent opacity variations due to $PH_3$. A coupled climate-photochemistry model that can provide self-consistent temperature-pressure profiles is required to probe this scenario.

#### 4.1.3 Phosphine Surface Fluxes Required for Detection

More critical than atmosphere abundances is the surface flux (i.e. the biological production rate) required for phosphine to accumulate to detectable abundance levels in anoxic atmospheres. This quantity plays a key role in determining the efficacy of $PH_3$ as a biosignature: if the presence of detectable levels of $PH_3$ in an atmosphere requires surface fluxes of $PH_3$ that are higher than that which a biosphere could plausible generate, then it is disfavored as a biosignature gas; if, on the other hand, $PH_3$ accumulates to detectable concentrations at fluxes within the range of plausible biological productivity, it is favored as a biosignature gas.

As phosphine moves up the atmosphere, its destruction rate and consequent mixing ratio change, due to the varying levels of radical concentrations and radiation at different altitudes. The dominant $PH_3$ reaction in $H_2$-dominated atmospheres is $PH_3$+H. The dominant reaction in $CO_2$-dominated atmospheres is $PH_3$+O. However, in high-$PH_3$ atmospheres, H produced from $PH_3$ photolysis becomes an increasingly important sink for $PH_3$, even in $CO_2$-dominated atmospheres. $PH_3$ is unlikely to dissolve into water and condense into aerosols (as ammonia, hydrogen sulfide, and methanethiol are) (Glindemann *et al.* 2003), meaning rainout is not expected to be a sink.

We use our photochemical model to estimate the minimum surface production flux, $P_{PH_3}$, for the detectability of phosphine in transmission and emission for a range of planetary scenarios (see Table 3). We find that, for planets orbiting active M-dwarfs, $PH_3$ can build to concentrations detectable by transmission and emission spectroscopy if produced at the surface with rates of the order of $10^{11}$ $cm^{-2}$ $s^{-1}$. We note that $PH_3$ requires similar surface flux rates in $H_2$- and $CO_2$-dominant atmospheres to reach detectable abundance levels, even though those correspond to much lower $PH_3$ concentration requirements in $H_2$-rich atmospheres than in $CO_2$-rich atmospheres. We speculate that this occurs because UV penetrates deeper into the more transparent $H_2$-rich atmosphere, allowing more radical accumulation and more photolysis at depth (see Figure 4 of (Hu *et al.* 2012)).

| **Atmospheric Scenario** | **Required Mixing Ratio for Detection (in transmission and emission)** | **$P_{PH_3}$ [$cm^{-2}$ $s^{-1}$]** |
|---|---|---|
| $H_2$-rich planet, Sun-like star | 780 ppm (transmission) | $1 \times 10^{14}$ |
| $H_2$-rich planet, active M-dwarf | 5 ppb (transmission) | $1 \times 10^{10}$ |
| $H_2$-rich planet, active M-dwarf | 220 ppb (emission) | $1 \times 10^{11}$ |
| $H_2$-rich planet, active M-dwarf ($PH_3$ runaway) | *0.28% (transmission)* | *$9 \times 10^{11}$* |
| $CO_2$-rich planet, active M-dwarf | 310 ppm (transmission) | $3 \times 10^{11}$ |
| $CO_2$-rich planet, active M-dwarf | 15 ppm (emission) | $1 \times 10^{11}$ |
| $CO_2$-rich planet, active M-dwarf ($PH_3$ runaway) | *7.6% (transmission)* | *$1 \times 10^{12}$* |

**Table 3**: Phosphine mixing ratios needed for detection in transmission and emission for different atmospheric and stellar scenarios, as well as associated surface flux requirements ($P_{PH_3}$ [$cm^{-2}$ $s^{-1}$]). The values in red represent surface fluxes and associated atmospheric abundances where $PH_3$ would be able to approach detection but would require longer than 200 of observation (which is longer than our allowed limit for detectability). Values in italic correspond to atmospheric scenarios where $PH_3$ is at a runaway threshold (see Figure 5, and Sections 4.1.3 and 5.1). For comparison, the maximum recorded surface flux of $PH_3$ on Earth is $10^{14}$ $cm^{-2}$ $s^{-1}$ (above sewage plants, Devai *et al.* 1988), and the biological production of $CH_4$ on Earth corresponds to $1.2 \times 10^{11}$ $cm^{-2}$ $s^{-1}$ (Guzmán-Marmolejo and Segura 2015; Houghton 1995; Segura *et al.* 2005).

The phosphine surface fluxes required to generate the detectable levels of $PH_3$ are large when compared to global $PH_3$ emissions on Earth, but are comparable to the production rates of other major biosignature gases. For comparison, biological $CH_4$ and isoprene production on Earth are of the order of $10^{11}$ $cm^{-2}$ $s^{-1}$ (Guenther *et al.* 2006), where a significant proportion of modern terrestrial $CH_4$ production is anthropogenic (see, for example, (Houghton 1995) via (Segura *et al.* 2005) and (Guzmán-Marmolejo and Segura 2015). As a further comparison, the highest recorded surface flux of $PH_3$ on Earth is above sewage plants, where $PH_3$ production reaches $10^{14}$ $cm^{-2}$ $s^{-1}$ (Devai *et al.* 1988).

One of our most interesting findings is the existence of a critical phosphine surface production flux, past which $PH_3$ accumulation is efficient and the atmosphere transitions to a $PH_3$-rich state. We term this critical flux the "tipping point". This effect appears analogous to the "CO runaway" effect identified for early Earth (Kasting 2014; Kasting *et al.* 2014; Kasting *et al.* 1984; Kasting *et al.* 1983; Zahnle 1986). Past the tipping point, $PH_3$ production outpaces the ability of stellar NUV photons to destroy $PH_3$, whether via direct photolysis or via generation of radical species. In this runaway phase, $PH_3$ can accumulate to percent levels and pervade the atmosphere. In this case, our models show that $PH_3$ can be detected with observation times reaching under 10 hours (e.g., see Figure 5). The plausibility of such a $PH_3$ run-away effect is discussed in Section 5.1.

#### 4.1.4 Sensitivity Analysis to Temperature and Radiation Levels

Our approach prescribes a temperature-pressure profile, and consider only two possible stellar scenarios (Sun-like stars and active M-dwarfs). We conducted sensitivity analyses to assess the dependence of our results on these assumptions.

#### Sensitivity Analysis to Temperature

In our study, we assumed surface temperatures of 288K; in reality, worlds with a broad range of temperatures may be habitable. Temperature may affect phosphine concentrations through varying reaction rates of $PH_3$ with radicals, and through changing the concentration of $H_2O$ in the atmosphere, from which the radical species are largely derived. To test the sensitivity of our results to surface temperature, we calculated $PH_3$ profiles for $CO_2$- and $H_2$-rich atmospheres orbiting with detectable concentrations of $PH_3$ at 288K, for surface temperatures of 273K and 303K. For simplicity, in calculating the dry adiabatic evolution of the lower atmosphere, we approximated the specific heat capacities at constant pressure of $CO_2$ and $H_2$, by their values at 273K (Pierrehumbert 2010). We adjusted the surface mixing ratio of water vapor to 0.0036 and 0.026, corresponding to the vapor pressures at 273K and 303K at the same 60% humidity assumed at 288K. Figures 8 and 9 present the results of these sensitivity tests in the case of an $H_2$-dominated atmosphere orbiting an M-dwarf star.

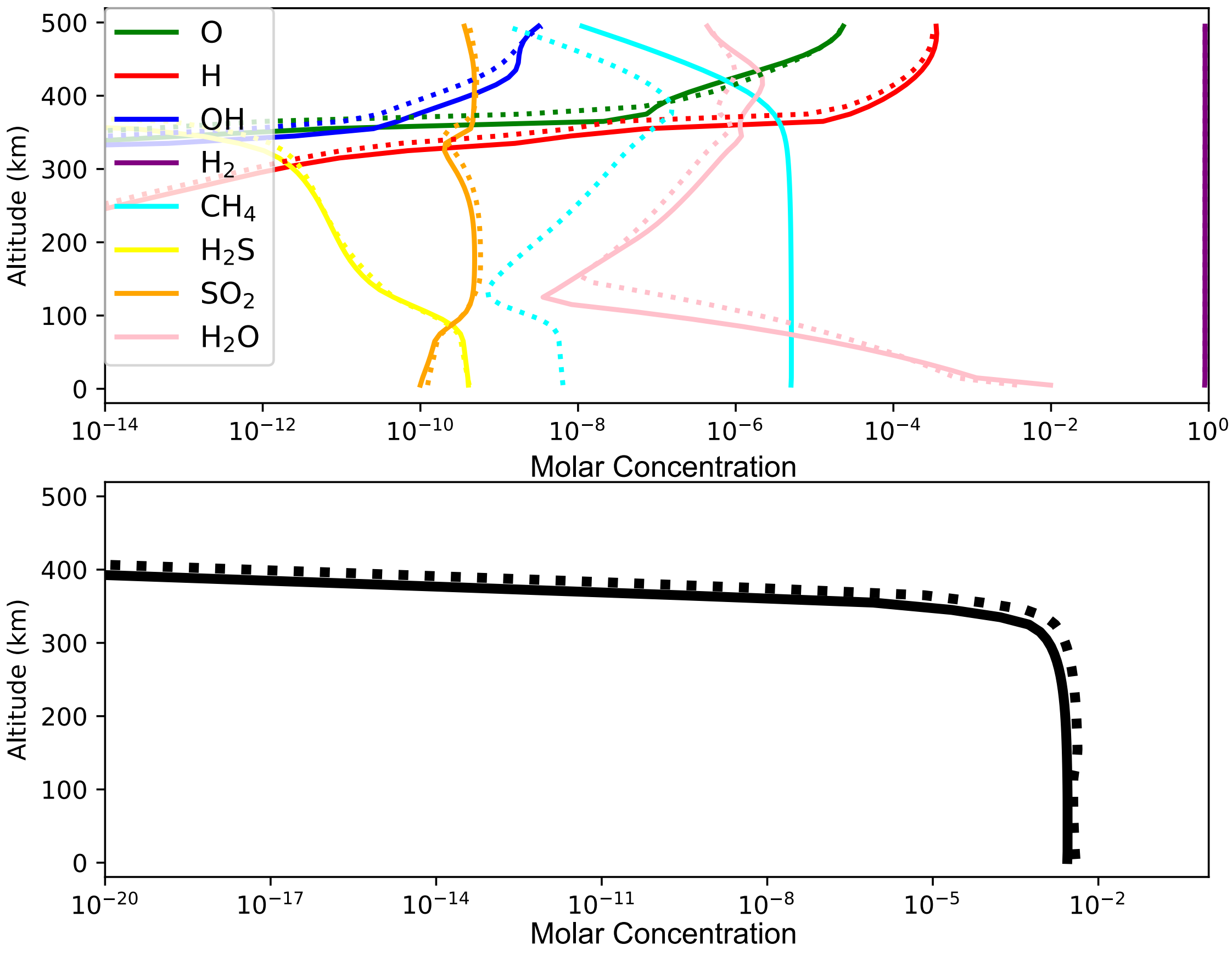

**Figure 8**: Distribution of abundances of atmospheric constituents (top panel) and phosphine (bottom panel) throughout the atmosphere of an $H_2$-rich planet orbiting an active M-dwarf. Vertical axes represent altitude in units of km, and horizontal axes represent molar concentration. Solid lines and dotted lines correspond to mixing ratios with surface temperatures of 288K and 273 K, respectively. When comparing low temperatures (273 K) to our standard 288 K models, the $PH_3$ mixing ratio remains mostly unchanged.

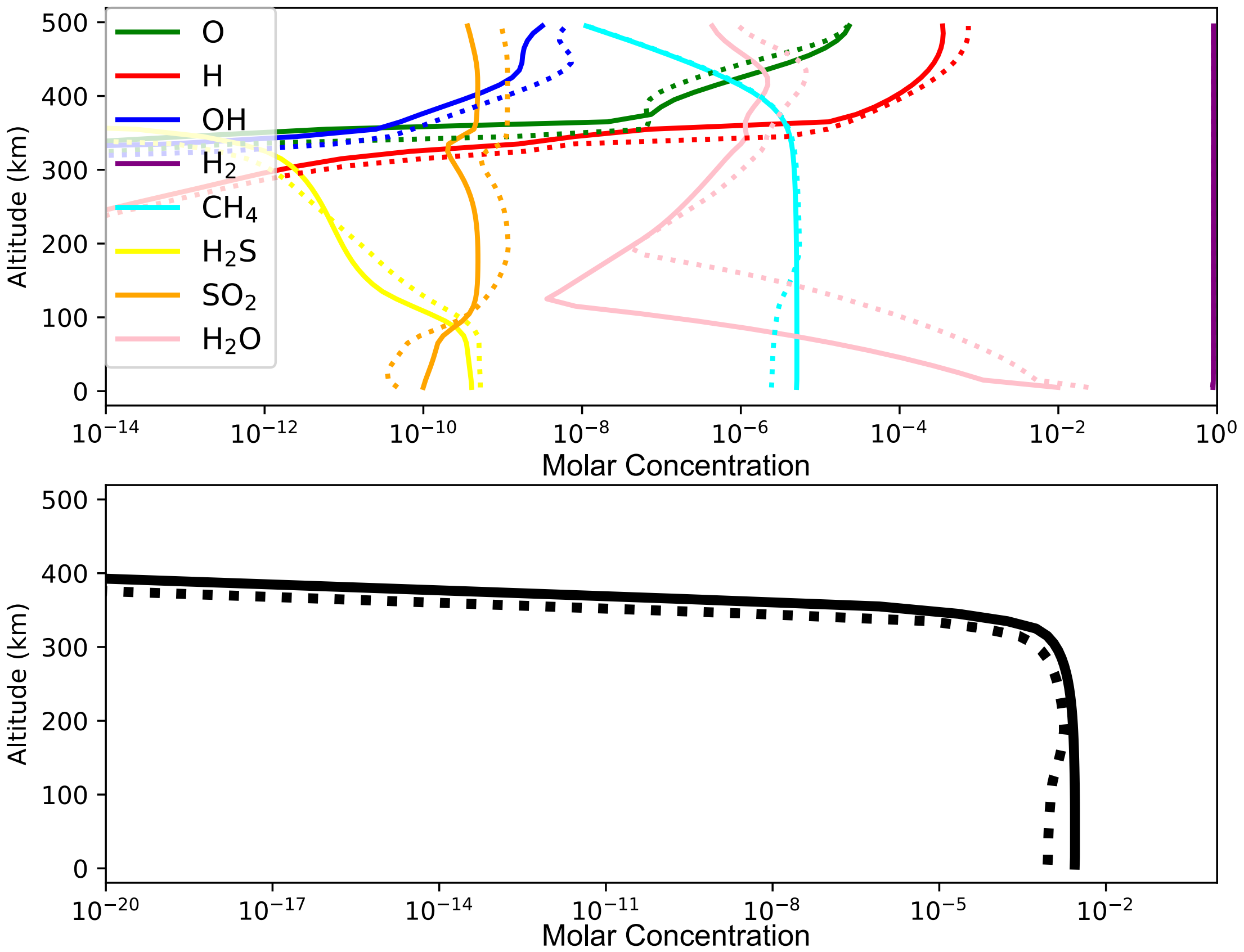

**Figure 9**: Distribution of abundances of atmospheric constituents (top panel) and phosphine (bottom panel) throughout the atmosphere of an $H_2$-rich planet orbiting an active M-dwarf. Vertical axes represent altitude in units of km, and horizontal axes represent molar concentration. Solid lines and dotted lines correspond to mixing ratios with surface temperatures of 288K and 303 K, respectively. When comparing high temperatures (303 K) to our standard 288 K models, the $PH_3$ mixing ratio remains mostly unchanged.

We find phosphine abundances to be weakly sensitive to surface temperature. For both $CO_2$- and $H_2$-dominated atmospheric scenarios, the total $PH_3$ column varies by a factor of ≤3 relative to the value at 288K across 273-303K, with the strongest variation occurring in $H_2$-dominated atmospheres. Such variations, while potentially significant for retrievals, do not affect our order-of-magnitude conclusions regarding the detectability of $PH_3$. We attribute this relatively modest variation of $PH_3$ column with temperature to the comparatively small variation of both the $PH_3$ radical reaction rates and the total water vapor column across this temperature range. Other atmospheric constituents, such as methane, show a much greater sensitivity to lower temperatures than $PH_3$. We are unsure why this is the case. One possibility is that $PH_3$ reaction rates are less sensitive to temperature changes than other atmospheric constituents (e.g. from 288 to 303 K, the rate constant for H + $CH_4$ increases by a factor of 2.3, whereas the rate constant for H + $PH_3$ increases by a factor of 1.16). Another possible explanation for $CH_4$ having a greater sensitivity to temperature than $PH_3$ is that $CH_4$ is primarily removed by OH (and therefore most sensitive to $H_2O$) while $PH_3$ is primarily removed by O and H (i.e. less sensitive to $H_2O$).

We crudely considered the potential impact of high phosphine abundances on the temperature profile of a planet. The greenhouse gas potential of $PH_3$ is not known (Bera *et al.* 2009) but it is plausible that a significant accumulation of $PH_3$ on an atmosphere would contribute to an increase in the global temperature since $PH_3$ is a strong IR absorber. To first-order, the change in surface temperature due to $PH_3$ can be estimated by calculating the surface temperature required to produce enough outgoing radiation to balance the arriving stellar radiation (see, e.g., (Pierrehumbert 2010). We executed this procedure for an atmosphere with and without $PH_3$. We estimate that, if $PH_3$ accumulates to the abundances required for its detection (see Section 4.1.3), $PH_3$ can lead to an increase of surface temperature between 10 and 30 K, depending on the atmospheric scenario. Further studies on the greenhouse gas potential of $PH_3$ are needed to fully explore the impact of its accumulation on the temperature profile of exoplanet atmospheres.

Overall, we conclude that our results are insensitive to variations in surface temperature of $\pm$15 K.

### Sensitivity to UV Irradiation

UV irradiation limits phosphine concentrations through direct photolysis and radical production. We considered the hypothesis that $PH_3$ would build to higher concentrations on a planet orbiting a star with low UV output, such as a quiet M-dwarf, as considered by (Domagal-Goldman *et al.* 2011). To test this hypothesis, we simulated $CO_2$-rich and $H_2$-rich planets orbiting a theoretical "quiet" M-dwarf. We constructed our quiet M-dwarf model by reducing the instellation at wavelengths <300 nm of our active M-dwarf case (corresponding to GJ1214) by a factor of 1000. This corresponds to ~100 times less UV than GJ 581, the quietest M-dwarf observed by the MUSCLES survey (France *et al.* 2016). A truly quiet M-dwarf may not exist, as practically all M-dwarfs observed to date have at least some chromospheric activity (France *et al.* 2013; France *et al.* 2016). Our quiet M-dwarf case may therefore be considered as a theoretical limiting case to study the effect of UV radiation on $PH_3$ buildup, with the understanding that this limiting case may not exist in reality. We nonetheless note that this limiting case is less extreme than photosphere-only limiting cases considered in past work (e.g., (Domagal-Goldman *et al.* 2011; Rugheimer *et al.* 2015; Seager *et al.* 2013b).

We find that, for the equivalent surface production rates, phosphine concentrations are two orders of magnitude higher on planets in the quiet M-dwarf cases compared to the active M-dwarf cases. Low UV emission favors buildup of $PH_3$ due to lower radical concentrations and photolysis rates. Consequently, as with other proposed biosignature gases, planets orbiting quiet M-dwarfs are the best targets for detecting biogenic $PH_3$ (Domagal-Goldman *et al.* 2011; Seager *et al.* 2013b; Segura *et al.* 2005). We also find that, in planets orbiting a quiet M-dwarf, $PH_3$ is able to enter a runaway phase with two orders of magnitude lower surface fluxes than those required in more active stars (Section 5.1).

Our overall main finding is that, because phosphine is easily destroyed either directly by UV or indirectly by UV-mediated creation of H, O, or OH radicals, a UV-poor environment is favorable for the detection of $PH_3$. This result is consistent with past work (Domagal-Goldman *et al.* 2011; Seager *et al.* 2013b; Segura *et al.* 2005). If there are no sufficiently quiet M-dwarf stars, we speculate that a UV-poor environment can be created by a UV-shield on the planet itself (e.g., (Wolf and Toon 2010)). Since $PH_3$ is readily destroyed in an $O_2$-rich environment, an ozone UV shield is unsuitable because other oxygen-containing radicals would destroy $PH_3$. However, elemental sulfur aerosols generated on planets with high volcanism and reducing atmospheres may provide such a UV-shield (Hu *et al.* 2013). Additionally, if $PH_3$ fluxes are high enough, they can overwhelm the supply of destructive UV photons and build up to higher concentrations (the "tipping point"). For more context and for a comparison with $CH_3Cl$, another proposed biosignature gas, see Sections 5.1 and 5.3.

## 4.2. Phosphine Spectral Distinguishability

Phosphine's spectral features can be easily distinguished from that of other gases expected to be main components of rocky planet atmospheres. Such gases include water vapor, methane, carbon dioxide, carbon monoxide, ammonia and hydrogen sulfide (see Figure 10). Ammonia might be present in hydrogen-rich atmospheres (Seager *et al.* 2013b).

The infrared spectrum of phosphine has three major features: 2.7 – 3.6 microns, 4-4.8 microns and 7.8 -11 microns, corresponding to polyad (P) numbers 3, 2 and 1, respectively (Sousa-Silva *et al.* 2013). The 2.7 – 3.6 microns region (P = 3) is dominated by a hot band and an overtone band, both associated with the symmetric bending mode of $PH_3$, and six additional combination bands. The 4-4.8 microns region (P=2) is dominated by both the fundamental symmetric and asymmetric stretching bands of the $PH_3$ molecule. The P=2 feature is also where several weaker, bending overtone and combination bands occur, combining with the fundamental bands to result in the strongest overall spectral feature for $PH_3$. When compared to water, methane, ammonia and hydrogen sulfide (but not $CO_2$) the P = 2 feature is uniquely attributable to $PH_3$ (Figure 10). In the 7.8-11 microns region (P=1), the fundamental symmetric and asymmetric bending modes, as well as hot bands, combine to produce a strong and broad absorption feature. The P=1 $PH_3$ feature overlaps with the ammonia spectrum but is easily distinguishable from the remaining molecules in this comparison (Figure 10).

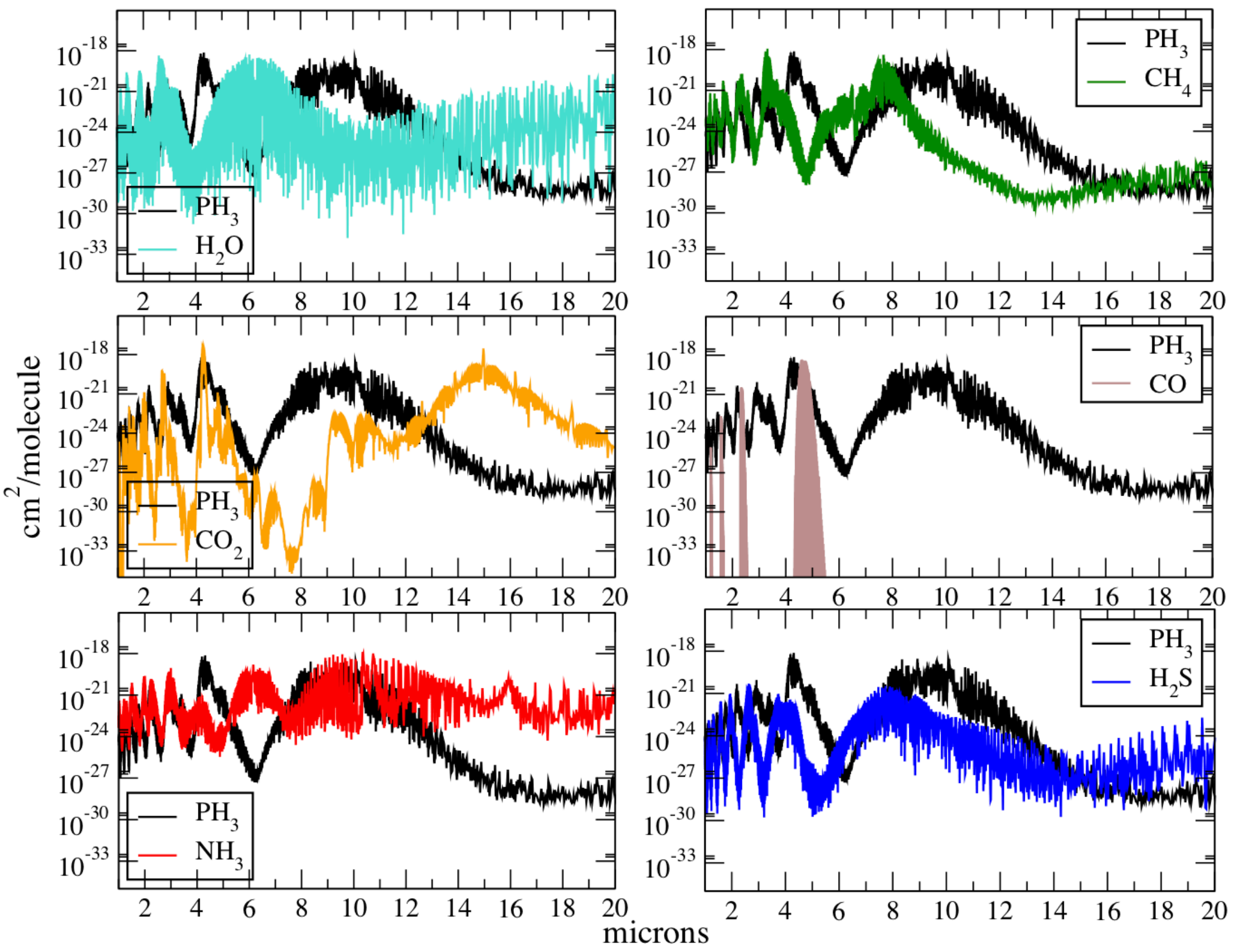

**Figure 10.** Comparison of the spectral cross-sections of phosphine with other molecular gases at room temperature. Intensity on y-axes in a log-scale with units of cm$^2$/molecule and wavelength represented on the x-axes in microns. All cross-sections are calculated at zero-pressure (i.e. Doppler-broadened lines only) using the procedure described by (Hill *et al.* 2013). $PH_3$, shown in black, is distinguishable from all compared molecules due to its strong bands in the 2.7 – 3.6 microns, 4-4.8 microns and 7.8 -11 microns regions.

For more detailed comparisons focusing on the 2.7 – 5 microns and 7.8 – 11.5 microns regions, see Appendix B.

The strongest band of phosphine, in the 4-4.8 microns region, is particularly salient when comparing $PH_3$ to all available spectra of volatile molecules (see Figure 11). However, the second strongest feature of $PH_3$, the broad band in the 7.8-11 microns region, is easily obscured by other gases as it absorbs in a heavily populated wavelength region, where many molecules have strong fundamental rovibrational modes.

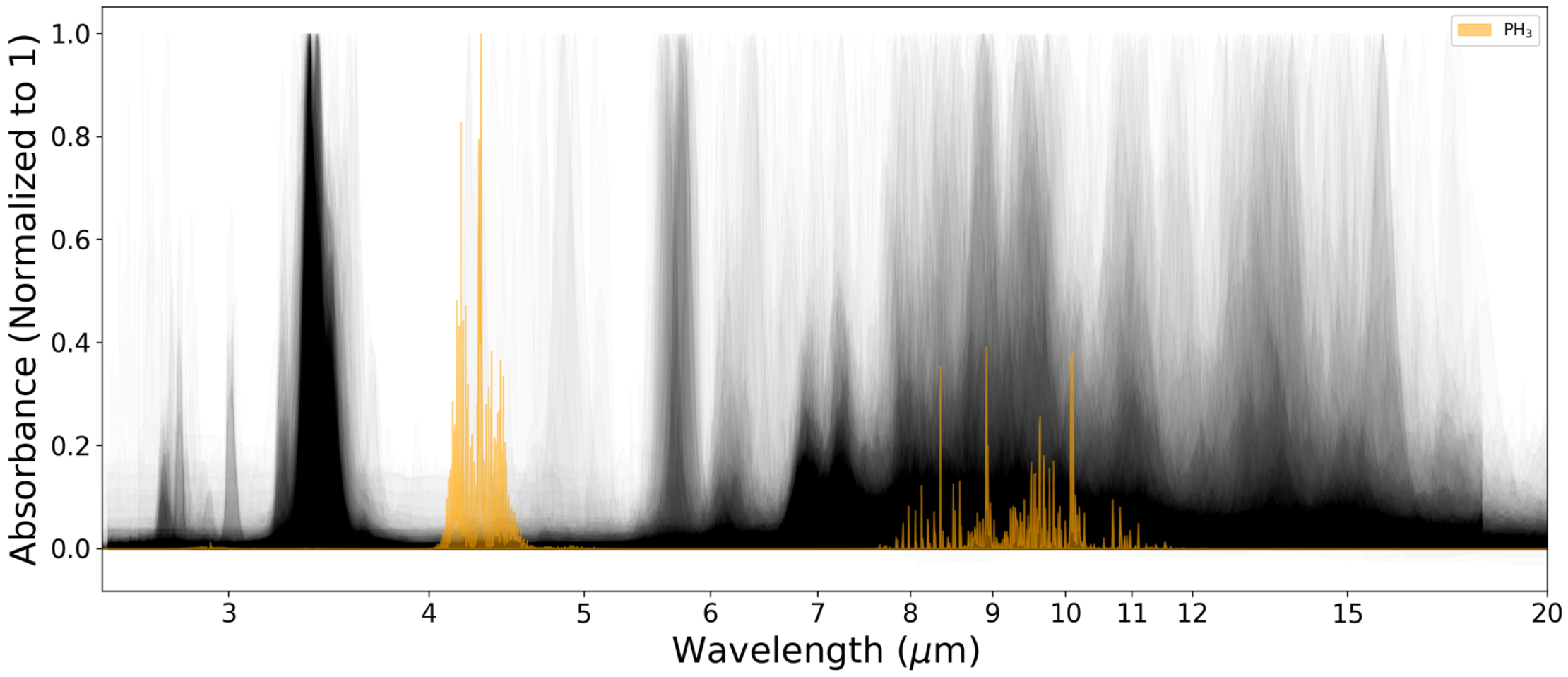

**Figure 11.** Comparison of the spectral cross-sections of phosphine (orange) with all the available cross-sections for molecules that are volatile at room temperature (Lemmon *et al.* 2010). Intensity on y-axes in a linear scale representing absorbance (normalized to 1) and wavelength represented on the x-axes in microns, with the spectral range constrained to 2.5 – 18.5 microns for fair comparison (many molecules have incomplete spectra beyond this region). Opacity for all molecules is plotted at 1% so that heavily populated regions are highlighted. All cross-sections are calculated with SEAS, using molecular inputs from NIST (Linstrom and Mallard 2001) and ExoMol (Tennyson *et al.* 2016). The strongest band of $PH_3$ (4.0-4.8 microns) is easily distinguishable from all other gases, but the broad band at 10 microns can become obscured by other molecules.

It is worth noting that, out of the 534 molecules for which there are available spectra, only a few dozen have been adequately measured or calculated, and consequently their spectra should be considered preliminary. Furthermore, there are thousands of volatile molecules which could contribute to an atmospheric spectra (Seager *et al.* 2016) for which there is no available spectra, so further studies are required to reveal the full extent of the spectral comparison highlighted in Figure 11 (Sousa-Silva *et al.* 2018; Sousa-Silva *et al.* 2019).

## 4.3. Phosphine False Positives

On Earth, the only significant amounts of phosphine found in the atmosphere are produced anthropogenically or biologically (Sections 2.1 and 2.2). The formation of $PH_3$ on temperate, rocky planets is thermodynamically disfavored, even in high-reducing environments, unlike the fermentative production of methane or hydrogen sulfide. In thermodynamic equilibrium, phosphorus can be conservatively expected to be found in the form of $PH_3$ only at $T > 800K$, and at $P > 0.1$ bar (Visscher *et al.* 2006), which is why $PH_3$ has been detected in Jupiter and Saturn, where these extreme temperatures occur (in the deep layers of the atmosphere). We also note that the critical temperature of water is 647 K so there are no surface conditions that favor both $PH_3$ production and allow for the presence of liquid water. Consequently, in a temperate, rocky planet, it is implausible that $PH_3$ can be produced without biological intervention, so its detection in such an environment is a promising indication of biological activity. We summarize below the potential false positive scenarios for $PH_3$ as a biosignature gas and their expected impact on the global concentrations of $PH_3$.

**Phosphite and phosphate disproportionation**: We considered the hypothesis that phosphine could be formed geochemically as a 'false positive' by reduction of phosphate or phosphite to $PH_3$. Phosphate is a dominant form of phosphorus on Earth. Phosphite is much less abundant but was detected in ground water and in mineral deposits (Han *et al.* 2013; Han *et al.* 2012; Yu *et al.* 2015) where it is likely to be the result of biological activity (Bains *et al.* 2019a). Phosphite was also postulated to be much more abundant on early, anoxic Earth (Herschy *et al.* 2018; Pasek 2008; Pasek *et al.* 2013). We calculated the Gibbs free energy of formation of $PH_3$ from both phosphate and phosphite under geochemical source conditions at neutral pH, for $T$ = 273 K and 413 K and $pH_2 = 10^{-6}$ bar and 1 bar. In all cases, the formation of $PH_3$ was thermodynamically disfavored; see Appendix C and D for details. We conclude that $PH_3$ formation from phosphate or phosphite is unlikely in the absence of a biological catalyst; for more details on the thermodynamic plausibility of the reduction of phosphites or phosphate into $PH_3$ see (Bains *et al.* 2019a).

Phosphite can disproportionate to phosphine at T > 323K and acidic pH (Bains *et al.* 2019a), raising the possibility that "black smoker" hydrothermal systems (T ≤ 678K, pH = 2-3, (Martin *et al.* 2008)) might generate phosphine (see Appendix D). Such systems do not dominate volcanic emission on Earth, leading us to propose they would be a negligible contributor on Earth-analog worlds. On the other hand, if a world had global, hot, acidic oceans (e.g., due to very high $pCO_2$), then the theoretical possibility of abiotic phosphine production exists, though likely only in the presence of high $H_2$ concentrations, very low pH and within a very hot temperature band (see Appendix D and Bains *et al.* 2019a). Given that these oceans would be unlikely to have pH values below 4 (carbonic acid has a pH of 3.6) and $PH_3$ formation is only favored at pHs closer to 2, we consider this scenario possible but implausible.

**Lightning:** We also considered the possibility of phosphine production by lightning. Lightning discharges even in highly reducing atmospheres produce only negligible amounts of reduced phosphorus species, including $PH_3$, and are very unlikely to provide high flux sources of $PH_3$ globally. A few studies have examined the production of reduced phosphorus species from phosphate as a result of simulated lightning discharges in laboratory conditions (Glindemann *et al.* 1999; Glindemann *et al.* 2004); only a very small fraction of the phosphorus was reduced to $PH_3$ through this process, even in highly reduced atmospheric conditions (Glindemann *et al.* 1999; Glindemann *et al.* 2004). Similarly, a mineral fulgurite - a glass resulting from lightning strikes - was also proposed as a potential source of $PH_3$ given that it could, in principle, contain reduced phosphorus species (Pasek and Block 2009) However, these sources are rare and localized; they would have minimal impact on a global scale. We are not aware of kinetically favored reactions that would promote the conversion of the thermodynamically favored phosphate to $PH_3$.

**Volcanism**: Phosphine is not known to be produced by volcanoes on Earth. Calculations on the production of $PH_3$ through vulcanism on a simulated anoxic early Earth showed that only trace amounts of $PH_3$ can be created through this avenue; the predicted maximum production rate is 102 tons per year (Holland 1984), which corresponds to ~$10^4$ $cm^{-2}$ $s^{-1}$. We note that the estimation of the maximum production of $PH_3$ through the volcanic processes reported by (Holland 1984) is made under the assumption of a highly reduced planet, which provides favorable conditions for $PH_3$ volcanic production. The volcanic production of $PH_3$ in other planetary scenarios is even more unlikely. We estimate that the maximum production of $PH_3$ by volcanoes in any planetary scenario, even $H_2$-rich atmospheres, is at least seven orders of magnitude lower than the surface fluxes required for detection (see Section 4.1.3).

**Exogenous delivery**: Finally, we considered the possibility of exogenous meteoritic delivery as a source of reduced phosphorus species that could lead to the abiotic production of phosphine. Reduced phosphorus species can be found in the meteoritic mineral schreibersite (Pech *et al.* 2011). Schreibersite is $(Fe,Ni)_3P$, which is present in iron/nickel meteorites (Geist *et al.* 2005); it is not present in stony or carbonaceous bodies. The current accretion rate of meteoritic material to the Earth is of the order of 20-70 kilotonnes/year (Peucker-Ehrenbrink 1996). We calculated the maximum $PH_3$ production from these sources as follows: considering that approximately 6% of the meteoritic material is iron/nickel (Emiliani 1992) and such meteorites contain an average of 0.25% phosphorus by weight (Geist *et al.* 2005), and working under the conservative assumption that the totality of the phosphorus content could be hydrolysed to $PH_3$, these meteors would deliver a maximum of ~10 tonnes of $PH_3$ to the Earth every year[5]. Therefore the contribution from meteoritic sources to the global average $PH_3$ production rates is still negligible. The above calculations are also in agreement with previous estimations of the phosphine production through meteoritic delivery, which were also found to be negligible (Holland 1984).

Overall, non-biological $PH_3$ formation is not favored on temperate rocky worlds, and no abiotic pathways can produce $PH_3$ with production rates necessary for its detection on habitable exoplanets. We therefore conclude that, in contrast to molecules like ammonia and methane, a *detection of $PH_3$ on a temperate exoplanet is likely to only be explained by the presence of life*.

## 5. Discussion

We find that phosphine is a promising marker for life if detected on a temperate exoplanet. On Earth $PH_3$ is naturally associated exclusively with anaerobic life, and is expected to not have any significant false positives for life on temperate exoplanets. Our models find that, if produced at sufficiently high surface fluxes, $PH_3$ can accumulate in planetary atmospheres to detectable abundances. Here we discuss the photochemical impact of high abundances of $PH_3$ in an exoplanet atmosphere (Section 5.1). We then describe the known limitations of our calculations (Section 5.2) and expand on alternative methods of detecting $PH_3$ in exoplanet atmospheres (Section 5.3). We summarize our findings in section 5.4.

---

[5] For comparison, 10.2 million tons per year of methane are produced from ruminants alone (Moss *et al.* 2000).

## 5.1. Phosphine "Tipping Point" and its Impact on the Atmosphere

Our models show that, with global phosphine surface fluxes comparable to those found locally in anoxic ecosystems on Earth, $PH_3$ can have a significant impact on planets with anoxic atmospheres. We calculate that $PH_3$ becomes detectable on anoxic planets where it is emitted at the surface with fluxes greater than $10^{11}$ $cm^{-2}$ $s^{-1}$. Consequently, $PH_3$ is accessible to remote detection only if it is a substantial product of the biosphere, emitted in quantities similar to $CH_4$ and isoprene on Earth (Guenther *et al.* 2006). $PH_3$ may be emitted as a product of primary metabolism, like $CH_4$, on warm, acidic worlds, or as a secondary metabolite, like isoprene (Bains *et al.* 2019a; Bains *et al.* 2019b).

A surprising result of our models was that, once the phosphine surface flux reaches a *tipping point* (e.g. > 9 x $10^{11}$ $cm^{-2}$ $s^{-1}$ for active M-dwarfs), $PH_3$ enters a runaway phase and begins to drastically change the atmosphere (see, e.g., Figure 5). This phase appears analogous to the CO runaway discovered for early Earth and reviewed in (Kasting *et al.* 2014). In this runaway phase, $PH_3$ production outpaces the ability of stellar NUV photons to destroy $PH_3$ (whether via direct photolysis or via generation of radical species), and modest increases in $PH_3$ flux lead to dramatic increases in $PH_3$ accumulation. With fluxes of $10^{12}$ $cm^{-2}$ $s^{-1}$ (10 times higher than the minimum required for detection) $PH_3$ approaches percent concentrations, and would be detectable with just less than 10 hours of observation (see Table 1 and Figure 12). In this runaway phase, $PH_3$ can affect the concentrations of other atmospheric constituents; for example, in $CO_2$-dominated atmospheres the $H_2$ concentration increases dramatically in $PH_3$ runaway, presumably from $H_2$ generated by $PH_3$ destruction. This raises the possibility that enhanced $H_2$ concentrations may be used to confirm $PH_3$ detections. In summary, $PH_3$ is readily detectable if it is produced at rates about an order of magnitude higher than methane and isoprene on Earth.

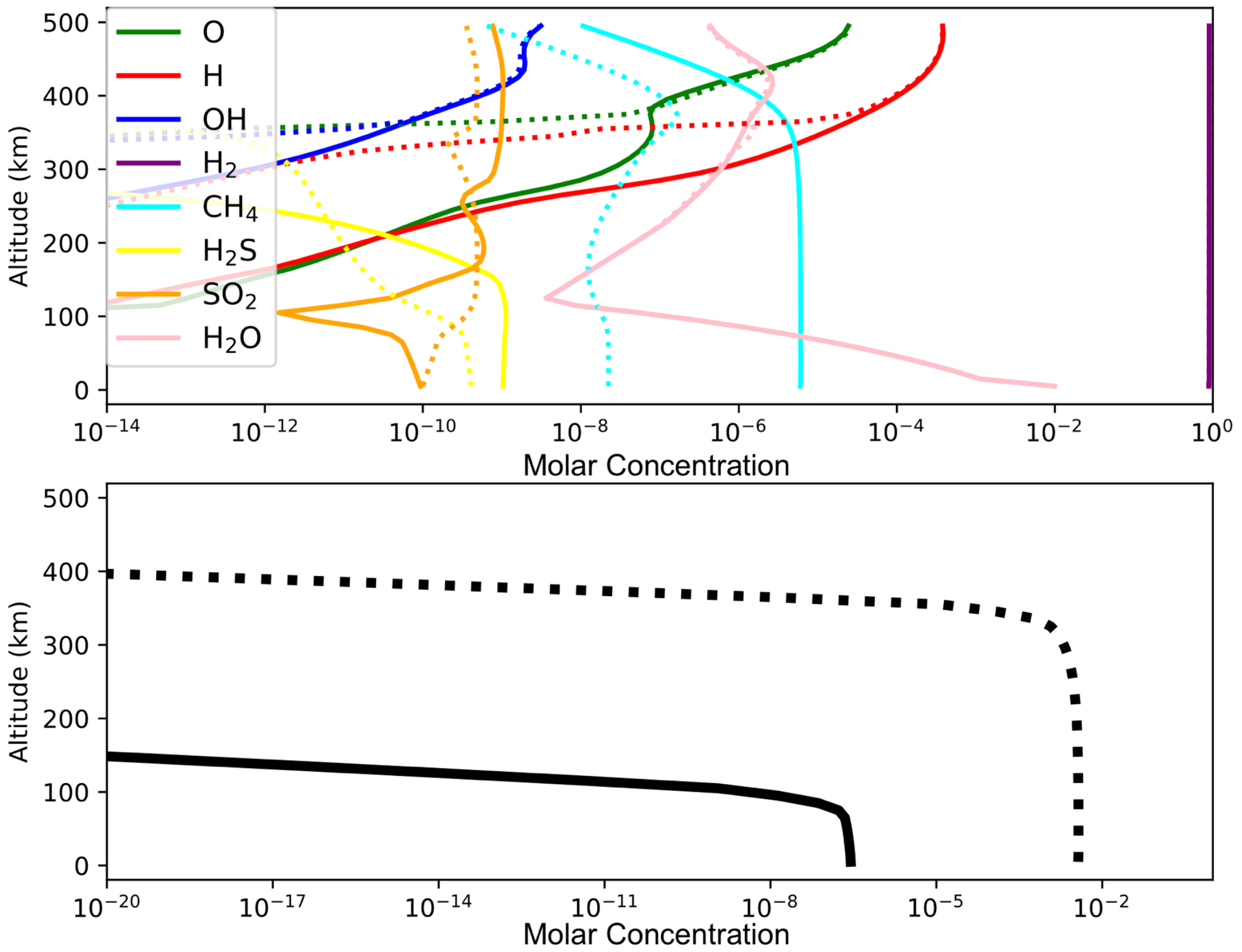

**Figure 12**. Distribution of abundances of major atmospheric constituents (top panel) and phosphine (bottom panel) throughout the atmosphere of an $H_2$-rich planet orbiting an active M-dwarf. Solid and dashes show molecular abundances immediately below and above the $PH_3$ runaway phase, respectively. The x-axis shows abundance concentrations, and the y-axis shows altitude in units of km. The scavenging effect of $PH_3$ leads to a decrease in O and OH and, to some extent, H radicals in the atmosphere. Consequently, both $PH_3$ and other trace gases (e.g. $H_2S$) are able to accumulate to larger abundances once $PH_3$ enters the runaway phase.

We considered whether the availability of phosphorus in the crust of a planet could be a limiting factor for the accumulation of phosphine in the atmosphere. An estimate of the total phosphorus within the Earth's crust shows that, if all the phosphorus was to be converted to $PH_3$, it would produce approximately twice as many $PH_3$ molecules as the total number of all molecules belonging to all gases present in the atmosphere of the modern Earth (Yaroshevsky 2006). We conclude that, in principle, the total mass of phosphorus in a planetary crust does not limit the development of a high-$PH_3$ atmosphere.

Phosphine can affect the spectrum of a rocky planet atmosphere, even at concentrations somewhat below detectability, by driving down radical concentrations due to its intense reactivity with these molecules, effectively becoming a scavenger in the atmosphere. This affects the concentrations of other atmospheric constituents, like methane. (Domagal-Goldman *et al.* 2011) described a comparable effect in models of organosulfur volatiles, which detectably altered the $CH_4$ and $C_2H_6$ abundances despite being themselves undetectable. (Domagal-Goldman *et al.* 2011) reported that elevated $C_2H_6/CH_4$ ratios could be diagnostic of high organosulfur flux, and hence a biosignature. Similarly, it may be possible to use the indirect effects of high $PH_3$ flux to infer the presence of $PH_3$ even if it is not directly detectable; more detailed measurements of the reaction kinematics of $PH_3$ and its byproducts, as well as more sophisticated atmospheric modelling is required to explore this possibility.

## 5.2. Model Limitations and Assumptions

We have assessed the potential of phosphine as a biosignature gas using a set of sophisticated photochemical and radiative transfer models. Nonetheless, given the complexity of simulating both atmospheric composition and subsequent spectral observations, many approximations and assumptions were made. Below is a brief discussion of the major limitations of the work performed here.

**Clouds**: We assumed cloudless skies for the simulation of exoplanet transmission spectra. To estimate how clouds might affect our results, we re-ran our models for phosphine abundances in the detectable range considering cloud decks at various altitudes, with coverage ranging from 10% to 100%. As expected, the detectability of $PH_3$ is reduced with the introduction of cloud coverage; for example, for planets with an $H_2$-dominated atmosphere orbiting an active M-dwarf, and with $PH_3$ surface fluxes of $10^{11}$ $cm^{-2}$ $s^{-1}$, the detectability of $PH_3$ reduced by up to a factor of 10 (from a cloudless model to a model with full cloud coverage at 100 Pa). However, once $PH_3$ reaches the tipping point (see Section 5.1), it becomes sufficiently abundant in the upper troposphere to be mostly unaffected by the presence of clouds.

**Reaction networks and haze formation**: In this work, we have focused on the reactivity of phosphine with the dominant radical species O, H, and OH. The photochemistry of $PH_3$ with radicals originating from other, more exotic, atmospheric species, though likely to be small, is insufficiently studied. For example, photochemistry of $PH_3$ and hydrocarbons, through UV-radiation, could lead to the formation of complicated alkyl-phosphines (Guillemin *et al.* 1997; Guillemin *et al.* 1995), and in consequence increase the probability of hazes. Inclusion of these reactions would increase $PH_3$ destruction rates and hence increase the required surface fluxes for detection; however, since these species are not expected to be dominant radicals, the effect of their inclusion would be minor and should not affect our results. In contrast to the formation of hydrocarbon and sulfur-based hazes that have previously been thoroughly addressed (Arney *et al.* 2017; Domagal-Goldman *et al.* 2011), there is very little work on phosphorus-based hazes. The formation of such hazes is possible in theory (Guillemin *et al.* 1997; Guillemin *et al.* 1995; Pasek *et al.* 2011), and early lab experiments implied the possibility of formation of such organophosphine hazes in planetary atmospheres, but further studies are needed to properly address the plausibility and impact of organophosphine haze formation and its associated potential as a $PH_3$ sink. Our photochemical model will continue to update whenever we are able to expand our reaction networks.

We have neglected the formation of organic hazes in this work. There is evidence that such hydrocarbon hazes occurred on Earth, due to transient high levels of methane (Izon *et al.* 2017; Zahnle *et al.* 2019). Organic hazes are predicted to form at $[CH_4]/[CO_2]>0.12$ for M-dwarf stars, and $[CH_4]/[CO_2]>=0.2$ for Sun-like stars (Arney *et al.* 2016; Arney *et al.* 2017). A methanogenic biosphere producing high $CH_4$ fluxes is required to generate such high ratios; in our work, we have not considered such a biosphere. The net effect of hazes would be to facilitate phosphine buildup through attenuation of photolytic UV. However, these same hydrocarbon hazes would also cloak some of the $PH_3$ spectral features, though primarily not in the wavelength bands where $PH_3$ is a strong absorber.

We did not include in our models the recombination of phosphine via $PH_2$ + H → $PH_3$. The rate constant for this reaction is $1.1 \times 10^{-10}$ cm$^3$ s$^{-1}$ at 288K (Kaye and Strobel 1984). The rate constant for H attack on $PH_3$ is $3.3\times10^{-12}$ cm$^3$ s$^{-1}$ at 288K (Arthur and Cooper 1997). Consequently, if $[PH_2]/[PH_3] \geq (3.3\times10^{-12}$ cm$^3$ s$^{-1})/(1.1\times10^{-10}$ cm$^3$ s$^{-1})=0.03$, then reactions with H can reform $PH_3$ as fast as it is destroyed by H-attack, and substantially lower the $PH_3$ surface fluxes that are required for $PH_3$ to accumulate in the atmosphere. Detailed photochemical modeling is required to constrain whether such high $[PH_2]/[PH_3]$ is plausible, or whether other sinks will suppress $[PH_2]$. We note that, in models of Saturn's atmosphere, $[PH_2]/[PH_3] << 0.03$ (Kaye and Strobel 1984); if the atmospheres of terrestrial $H_2$-dominated exoplanets behave similarly, this mechanism will not be able to significantly replenish $PH_3$.

Finally, we note that our results reflect the prediction of atmospheric models that anoxic atmospheres should have much higher mean radical concentrations than modern Earth (Hu *et al.* 2012). Due to their reactivity, one might expect concentrations of radicals to be suppressed even in anoxic atmospheres, as OH is on Earth. While our reaction networks include the known relevant atmospheric chemistry, it is possible that there are chemical reactions which are relevant to anoxic, temperate terrestrial planets, but which have not been considered in the context of the Solar System and hence are not included in the reaction compendia we use in our model, e.g. the NIST database (Linstrom and Mallard 2001) and the JPL compendium (Sander *et al.* 2011). Due to the incompleteness of our reaction network, it is likely that, when phosphine is destroyed, additional radicals are created that are not considered in our models. While $PH_3$ is a trace gas, this omission should have negligible consequences, as the calculation of the infinite series due to radicals tends to converge quickly. However, when $PH_3$ approaches the runaway phase and becomes the dominant radical sink the atmosphere, our scenarios become a low-radical regime. In reality, the radical production would not stop, and the transition to a runaway scenario may be slower than we predict. If radical concentrations were overestimated in our model, then $PH_3$ can build to detectable levels with lower surface fluxes. If, at high concentrations of $PH_3$, the intermediate radical production from its destruction has been underestimated, then $PH_3$ can only enter a runaway phase with higher surface fluxes than those calculated here. Detailed studies of the chemical reaction networks of anoxic planets are required to explore these possibilities.

**Prescribed temperature-pressure profiles**: Our photochemical model uses prescribed temperature-pressure profiles (see Appendix A) that are isothermal above the stratosphere. In reality, this is an over-simplification; one of the consequences of assuming there is no temperature inversion at high altitudes is the underestimation of the detectability of phosphine in our simulated emission spectra. We also did not couple the potential heating effect from $PH_3$ dissociation in our photochemical model, so we performed a sensitivity analysis to small changes in temperature ($\pm$15 K) and found that our main conclusions remain unchanged. However, if $PH_3$ fluxes exceed the "tipping point" and $PH_3$ enters a runaway phase, it may be possible for temperatures to increase beyond the maximum 303 K we consider here (see Section 4.1.4). In the most extreme scenario, a $PH_3$ runaway might trigger a runaway greenhouse state, potentially rendering the planet uninhabitable. A coupled climate-photochemistry model is required to thoroughly investigate this scenario.

**Phosphine sinks**: Our photochemistry model assumes a dry deposition velocity of 0 for phosphine, i.e. no consumption of $PH_3$ by surficial geochemistry or biology. Apart from its efficient oxidation by atmospheric components there are no other known significant $PH_3$ sinks on Earth. It is possible however that other planets may have $PH_3$ deposition pathways that we cannot account for. For example, given the opportunistic nature of biology, and the fact that, at least on Earth, phosphorus is a growth-limiting nutrient, life might use any excess of atmospheric $PH_3$ as source of phosphorus. It is plausible that anaerobic life on other planets will not just produce $PH_3$ but also reabsorb it from the atmosphere; in these scenarios, biology would slow down the $PH_3$ accumulation in the atmosphere leading to a dampening, and possible avoidance, of a runaway $PH_3$ effect. In extreme cases, for example when biological production of $PH_3$ equals to its reabsorption, life's recycling of biogenic $PH_3$ might entirely prevent its accumulation in the atmosphere of an exoplanet.

To estimate the effect of a potential sink for phosphine, we tested the variability of our results to a non-zero deposition velocity for $PH_3$. Deposition velocity tests show that, for a comparable deposition rate to CO and $O_2$ ($10^{-4}$ $cm^{-1}$ (Harman *et al.* 2015) Harman et al 2015), concentrations of $PH_3$ varied by a factor of <2, which is not enough to affect detectability. We note this is a conservative estimate given that the only plausible $PH_3$ sinks are biological.

## 5.3. Alternative Detection Methods Beyond JWST

For the phosphine detectability calculations in this work we have considered observations from a JWST-like telescope, with a 6.5 m diameter telescope mirror operating within 50% of the shot noise limit and a quantum efficiency between 20-25%. The integration time is assumed to be under 200 hours for all atmospheric scenarios. For comparison, the cryogenic lifetime of JWST is 5 years, which is equivalent to an integration time of 100 hours for a planet orbiting the habitable zone of an M-dwarf star. The spectral resolution of JWST is R = 100 at 1 - 5 μm and R = 160 at 5 - 12 μm, which is more than necessary for distinguishing between $PH_3$ and other dominant gases in the atmosphere (see Section 4.2).

We considered the possibility of detecting phosphine in anoxic atmospheres using alternative telescopes to JWST. Missions such as TPF-I (Lawson *et al.* 2008), Darwin (Fridlund 2000), OST (Battersby *et al.* 2018), HabEx (Gaudi *et al.* 2018), LUVOIR (Roberge 2019), and the 30-meter class of ground telescopes (Johns *et al.* 2012; Skidmore *et al.* 2015; Tamai and Spyromilio 2014) could also be able to characterize atmospheres of temperate planets in wavelength regions where $PH_3$ is spectrally active.

TPF-I was intended to be a nulling interferometer with four 4-m diameter telescopes formation flying with a baseline range of 40-100 m and operating at 6.5 to 18 microns with a spectral resolution of 25-50. Darwin was planned as set of 3-4 m diameter telescopes flying in a nulling interferometer configuration. For both instruments, only the nearest (~4 parsec) M dwarf star habitable zones would be accessible. Winters *et al.* 2019 estimates that there are only 22 M-dwarf stars candidates that could have planets within their habitable zone suitable for atmospheric characterization. TPF-I and Darwin are currently cancelled, but similar telescopes (e.g., the proposed Large Interferometer For Exoplanets, or LIFE (Quanz *et al*. 2019)) may one day be commissioned that perform in a similar interferometer formation and could have the capability to identify phosphine on temperate exoplanets orbiting M-dwarfs.

The Origins Space Telescope (OST) (Battersby *et al.* 2018) has a large wavelength coverage (2.8–20 microns), and will be able to provide atmospheric spectra for planets orbiting K- and M-dwarf stars, through transmission and secondary eclipse observations. OST will have the sensitivity and coverage to detect many spectral signatures of potential biosignature gases, including phosphine.

The Habitable Exoplanet Imaging Mission (HabEx) (Gaudi *et al.* 2018) and the Large UV Optical Infrared Surveyor (LUVOIR) (Roberge 2019) are planned as powerful telescopes that could launch in the coming decades and are focused on the detection of potential biosignature gases. Both cover a wavelength region where phosphine is spectrally active (UV to near-IR), though not a particularly strong absorber (1.28 to 1.79 microns (Sousa-Silva *et al*. 2014)). We have not modelled atmospheres with $PH_3$ in an HabEx/LUVOIR scenario but these telescopes' high contrast spectroscopy may allow for some of the high-frequency combination and hot bands of $PH_3$ to be detected.

Future 30-meter ground-based telescope will be limited by the Earth's observing windows, but those capable of M- and N-band spectroscopy (4.5 - 5.0 and 7.5 - 14.5 microns, respectively) would be favorable for the detection of phosphine, given its strong feature at 4-4.8 microns, and its broad band centered at 10 microns (see Section 4.2).

## 5.4. Phosphine as a Biosignature Gas

An ideal biosignature gas lacks abiotic false positives, has uniquely identifiable spectral features, and is unreactive enough to build up to detectable concentrations in exoplanet atmospheres. Phosphine fulfills the first two criteria: $PH_3$ is only known to be associated with life and geochemical false positives for $PH_3$ generation are highly unlikely (Bains *et al.* 2017; Bains *et al.* 2019a); $PH_3$ possesses three strong features in the 2.7 – 3.6 microns, 4-4.8 microns and 7.8 -11 microns regions that are distinguishable from common outgassed species that may be present in terrestrial exoplanet atmospheres, such as $CO_2$, $H_2O$, CO, $CH_4$, $NH_3$, and $H_2S$.

The greatest challenge to the detectability of phosphine at low surface fluxes is its reactivity to radicals, and its vulnerability to UV photolysis. In the most tractable observational scenario (planet orbiting an active M-dwarf), $PH_3$ must be emitted at a rate of $10^{11}$ cm$^{-2}$ s$^{-1}$ to build to levels detectable by transmission or thermal emission spectroscopy (e.g., using JWST). The required $PH_3$ production rates for detection are two orders of magnitude lower for planets orbiting a hypothetical "quiet" M-dwarf with extremely low levels of chromospheric activity; this latter scenario is likely unrealistic and corresponds to an extreme lower limit on the required $PH_3$ flux. An alternative path for a UV-poor surface environment would be a planet with a UV shield, or possibly intense hazes, though hazes might also inhibit a $PH_3$ detection. Finally, our models suggest that, at high but plausible surface fluxes ($10^{12}$-$10^{14}$ cm$^{-2}$ s$^{-1}$, depending on the planetary scenario), $PH_3$ is able to exhaust the supply of M-dwarf NUV photons and enter a "runaway" phase. If such a runaway effect occurs, $PH_3$ becomes easily detectable, but also protects other trace gases from destruction by radicals and rapidly changes the overall composition of the planetary atmosphere.

We compare phosphine to another proposed biosignature gas: methyl chloride, or $CH_3Cl$ (Segura *et al.* 2005). The surface flux required for $CH_3Cl$ buildup to detectable levels is $5\times10^{7}$ – $3\times10^{11}$ cm$^{-2}$ s$^{-1}$ under the same atmospheric scenarios we consider here, or up to four orders of magnitude less than $PH_3$. This difference is rooted in the lower reactivity of $CH_3Cl$ with radicals; its reaction rates with O and H are 4-6 orders of magnitude lower than $PH_3$ (Seager *et al.* 2013b). We conclude that it is more difficult for $PH_3$ to build to detectable concentrations than $CH_3Cl$ because of its higher reactivity with radicals. As a further comparison, the modern methane surface flux is 535 Tg year$^{-1}$ = $1.2\times10^{11}$ cm$^{-2}$ s$^{-1}$, meaning that for $PH_3$ to build to detectable levels in planets orbiting active stars, it must be emitted at a rate comparable to that of methane on Earth ((Houghton 1995) via (Segura *et al.* 2005) and (Guzmán-Marmolejo and Segura 2015)).

For a final comparison we consider the highest concentrated fluxes of phosphine on Earth. On Earth, detections of $PH_3$ in biogas from sewage plants can reach very high levels, e.g. $10^{14}$ cm$^{-2}$ s$^{-1}$ (Devai *et al.* 1988). It is therefore plausible that a complex anaerobic biosphere, not dissimilar from those found in sludges on Earth, could achieve global $PH_3$ production levels comparable to the production rates found locally in isolated environments on Earth. In such anoxic environments, $PH_3$ would be detectable in transmission and emission for planets orbiting M-dwarfs (see Table 1). In other words, we can imagine, for example, a planet like Earth in the early Carboniferous period (~318 Mya), but without $O_2$ in the atmosphere: a very wet, anoxic, 'tropical paradise' from pole to pole. Such a planet could potentially produce tremendous amounts of $PH_3$ by its rich anaerobic biosphere.

Phosphine's relevance as a biosignature gas depends on its production by life, on its possible geochemical false positives, and on its detectability. Prospects for the buildup of $PH_3$ to detectable levels are uncertain with near-future telescopes. However, the wavelength regions where $PH_3$ is spectrally active are similar to those of other atmospherically important molecules, e.g. $H_2O$ and $CH_4$. Consequently, searches for $PH_3$ can be carried out at no additional observational cost to searches for other molecular species relevant to characterizing exoplanet habitability. Ultimately, if detected on a temperate planet, $PH_3$ is an extremely promising biosignature gas, since its lack of high-flux false positives would be a strong reason to hypothesize production by life.

## Acknowledgements

We thank the MIT BOSE Fellow program and the Change Happens Foundation for partial funding of this work. We thank Elisabeth Matthews, Thomas Evans, Julien de Wit, and Jason Dittmann for their advice on detectability metrics. We also thank Antonio P. Silva, Fionnuala Cavanagh, Catherine Wilka, Sarah Ballard, Sarah Rugheimer, Jennifer Burt, Daniel Koll, Susan Solomon, Andrew Babbin, Tiffany Kataria, Antonio Silva and Christopher Shea for their useful discussions and contributions. Finally, we would like to thank our two reviewers, whose contributions significantly improved this manuscript. This research was supported in part by a grant from the Simons Foundation (SCOL; grant # 495062 to S.R.), and carried out in part at the Jet Propulsion Laboratory, California Institute of Technology, under a contract with the National Aeronautics and Space Administration.

## Appendices

### A: Atmospheric Mixing Ratios and Temperature-Pressure Profiles

Figure A-1 shows the molecular mixing ratio profiles used to simulate atmospheric spectra for the planets with $CO_2$- and $H_2$-rich atmospheres orbiting active M-dwarfs, quiet M-dwarfs and Sun-like stars. To establish the effect of adding phosphine to the atmosphere, we ran the photochemical models with the starting mixing ratios from Figure A-1 considering surface fluxes of PH3 ranging from $10^3$ cm$^{-2}$ s$^{-1}$ to $10^{14}$ cm$^{-2}$ s$^{-1}$.

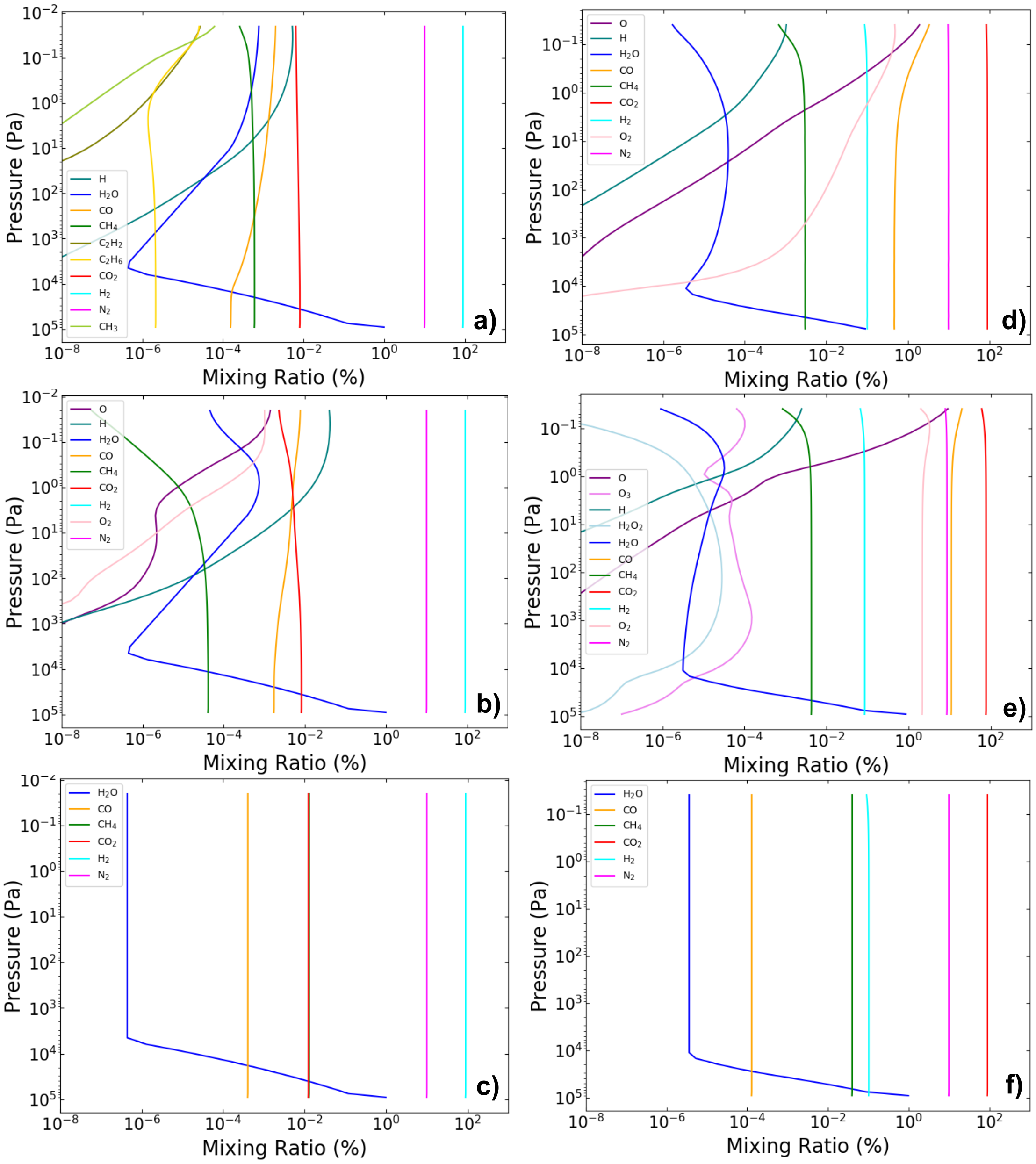

**Figure A-1**: Mixing ratio profiles of simulated Earth-sized planets with the following parameters: a) $H_2$-dominated atmosphere on a planet orbiting an Sun-like star; b) $H_2$-rich atmosphere on a planet orbiting an active M-dwarf; c) $H_2$-rich atmosphere on a planet orbiting a quiet M-dwarf; c) $CO_2$-rich atmosphere on a planet orbiting a Sun-like star; d) $CO_2$-rich atmosphere on a planet orbiting an active M-dwarf; e) $CO_2$-rich atmosphere on a planet orbiting a quiet M-dwarf; e). Vertical axis represents pressure in units of Pa and horizontal axis shows the mixing ratio represented as a percentage of the total atmosphere. Figure partially adapted from (Hu *et al.* 2012) and (Seager *et al.* 2013b).

Figure A-2 shows the temperature-pressure profiles used to simulate the transmission and emission spectra for the planets with $H_2$- and $CO_2$-rich atmospheres.

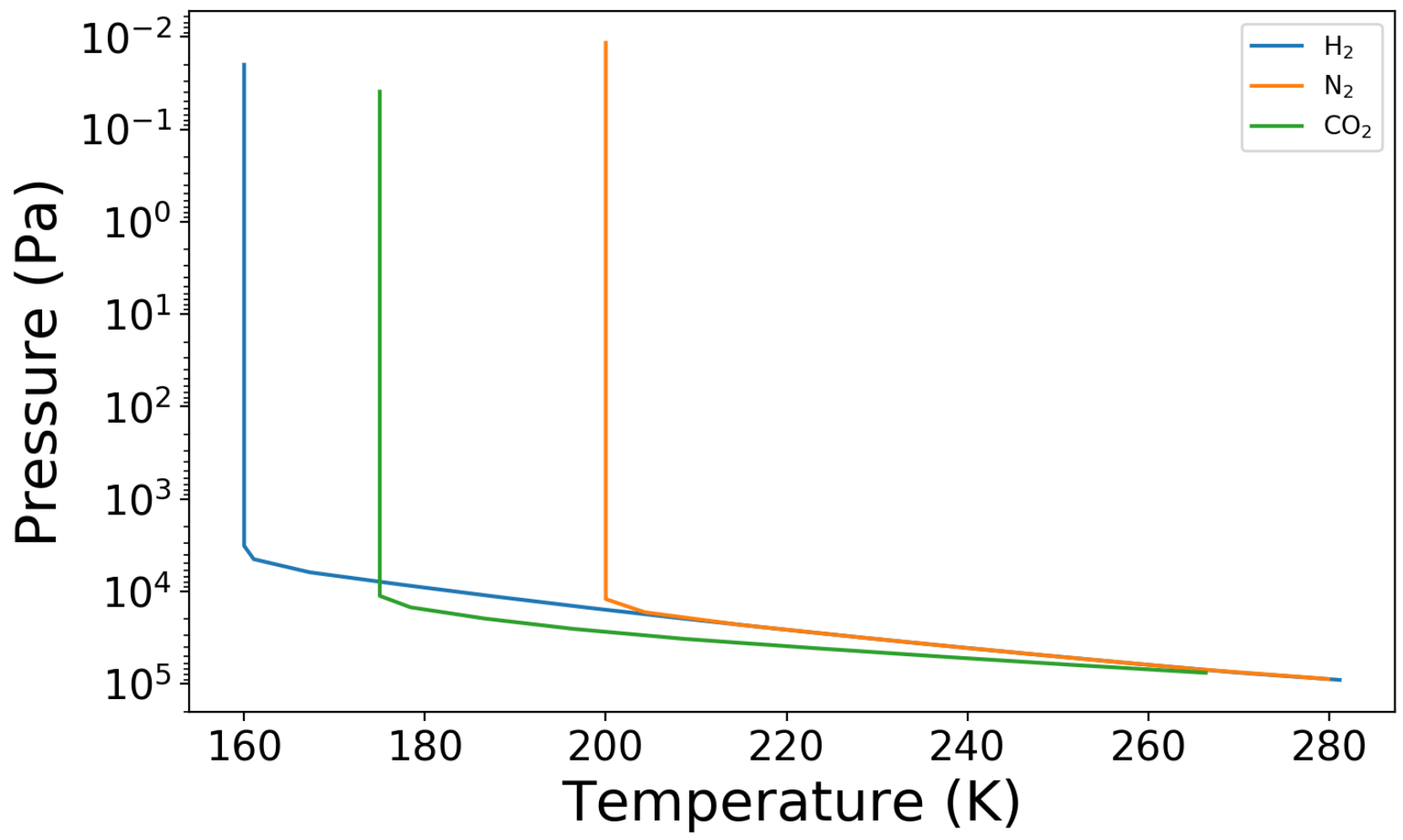

**Figure A-2**: Temperature-pressure profiles of the simulated massive super Earths ($M_p$ =10 $M_E$ and $R_p$ = 1.75 $R_E$), with $H_2$-, and $CO_2$-rich atmospheres ($N_2$-rich atmosphere shown for comparison as an intermediate reducing scenario). Vertical axis represents pressure in units of Pa and horizontal axis shows the temperature in units of K. Figure adapted from (Hu *et al.* 2012).

## B: Detailed Spectral Comparison of Phosphine with Other Atmospheric Components

The three strongest spectral features of phosphine occur at 2.7-3.6 microns, 4.0-4.8 microns and 7.8-11.5 microns. Figure B-1 and B-2 (2 - 6.5 microns, and 7.5 -11.8 microns, respectively) show a detailed comparison between the room temperature cross-sections of $PH_3$ and six common atmospheric molecules, namely $H_2O$, $CH_4$, $CO_2$, CO, $NH_3$ and $H_2S$.

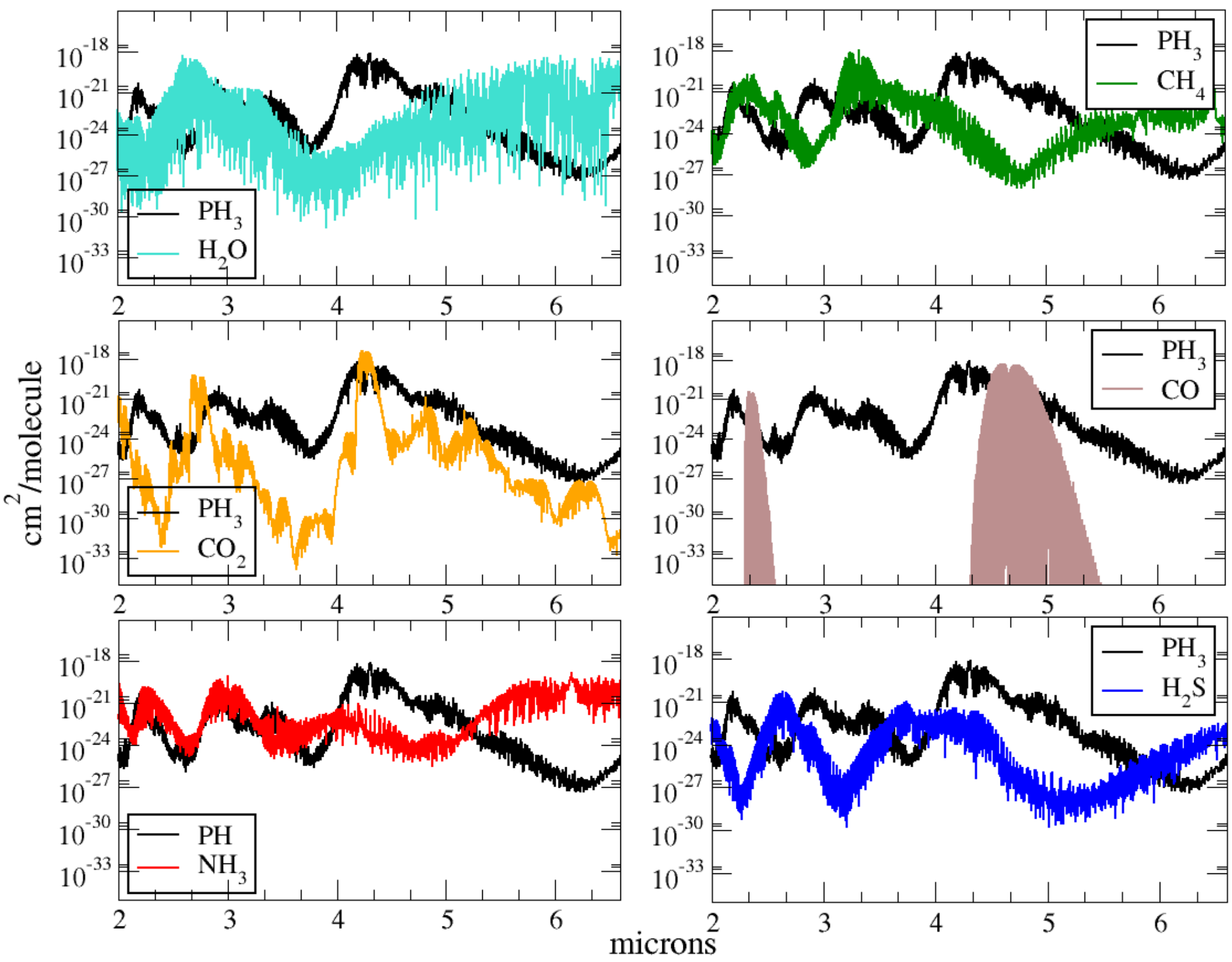

**Figure B-1.** Comparison of the spectral cross-sections of phosphine with other gases at room temperature, in the 2 – 6.5 microns region. Intensity on y-axes with units of cm$^2$/molecule and wavelength represented on the x-axes in microns. All cross-sections are calculated at zero-pressure (i.e. Doppler-broadened lines only) using the procedure described by (Hill *et al.* 2013). $PH_3$, shown in black, is distinguishable in this region from all compared molecules.

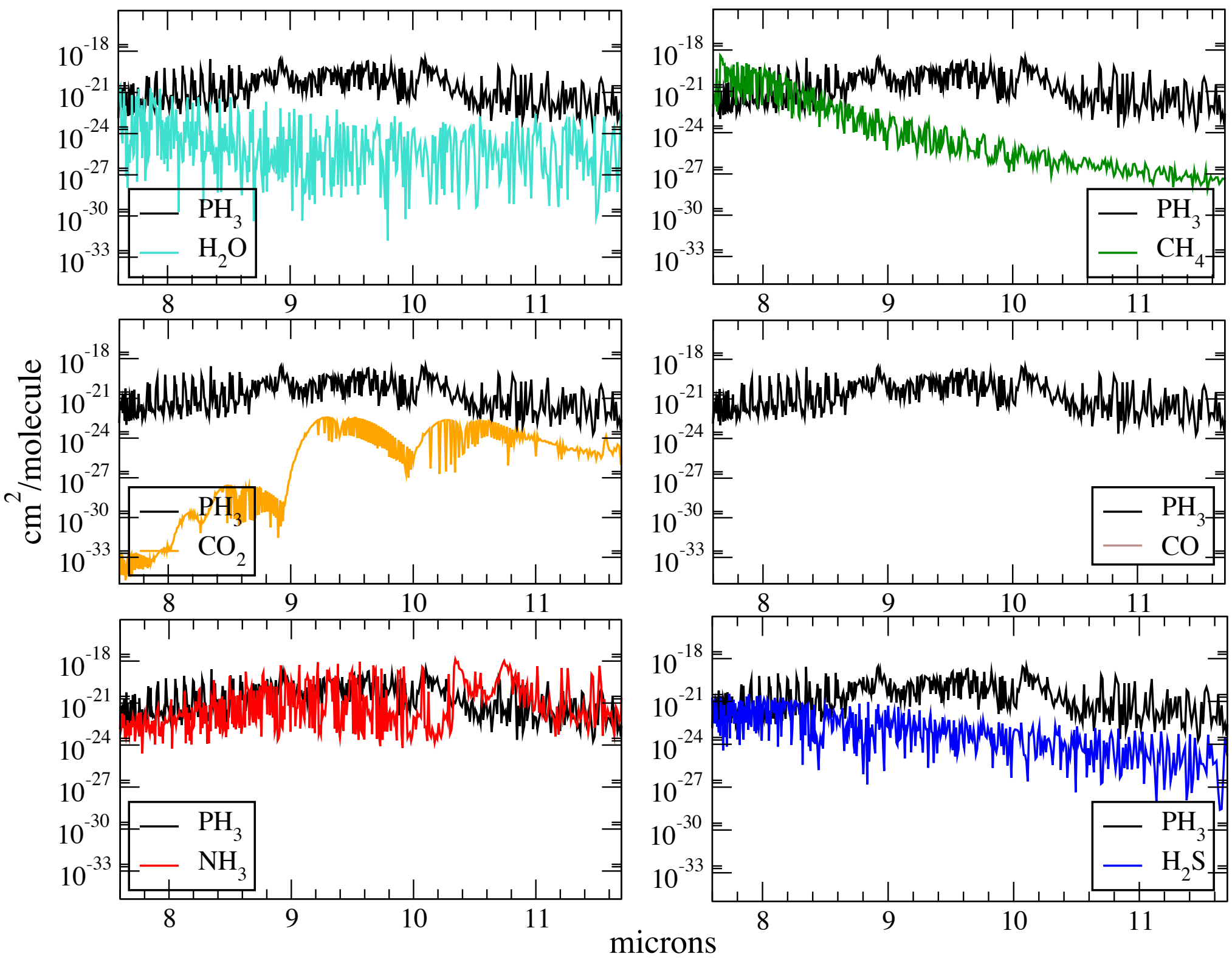

**Figure B-2.** Comparison of the spectral cross-sections of phosphine with other gases at room temperature, in the 7.6 – 11.7 microns region. Intensity on y-axes with units of $cm^2$/molecule and wavelength represented on the x-axes in microns. All cross-sections are calculated at zero-pressure (i.e. Doppler-broadened lines only) using the procedure described by (Hill *et al.* 2013). $PH_3$, shown in black, is distinguishable in this region from all compared molecules except ammonia. Note that CO is not known to absorb in this region.

## C: Thermodynamics of Abiotic Phosphine Synthesis at Circumneutral pH

We conduct a thermodynamic analysis to determine whether the formation of phosphine from phosphate and phosphite is thermodynamically favored at circumneutral pH. We calculate the free energy of formation of $PH_3$ in solution (and solid, in reduction of calcium phosphate) under geochemical source conditions and elemental composition similar to terrestrial planets for two temperature conditions (0°C and 140°C, or 273 K and 413 K) and two extreme $H_2$ abundance levels (Table C-1) on the basis of established thermodynamic values (Amend and Shock 2001; Barner and Scheurman 1978; Conrad *et al.* 1986; Fu *et al.* 2013; Linstrom and Mallard 2001). The free energy of reaction of formation of $PH_3$ is given by:

$$\Delta G = \Delta G^o + R.T.ln(Q),$$

where $\Delta G^o$ is the standard free energy of formation, $R$ is the gas constant, $T$ is the absolute temperature and $Q$ is the reaction quotient.

The resulting calculated free energy of formation of phosphine can be used as a proxy to estimate the likelihood of $PH_3$ formation as a geochemical false positive (Bains *et al.* 2017). The more positive the free energy of formation of $PH_3$ for a given reaction pathway the less likely it is to be produced through geochemical processes. We note that the estimation of likelihood of geochemical false positive scenarios for $PH_3$ is a part of the larger effort to estimate the possibilities for false positives for all biosignature gases (Bains *et al.* 2017). The calculated energy of formation of $PH_3$ is strongly positive for any of the proposed $PH_3$ formation reaction pathways, given the geochemical concentrations of gases, suggesting that if $PH_3$ is detected in an atmosphere of an exoplanet its source is not likely to be geochemical, increasing the probability for biological production (Table C-1).

**1) Oxidized atmosphere (low $H_2$ levels)**

| Possible geochemical $PH_3$ formation reaction | ΔG° (kJ/mol) | ΔG (kJ/mol) at plausible geochemical gas concentrations<br>$H_2 = 10^{-6}$ bar<br>$PH_3 = 10^{-6}$ bar<br>$CH_4 = 10^{-6}$ bar<br>$CO_2 = 0.01$ bar<br>phosphate = $10^{-5}$ M<br>phosphite = $10^{-6}$ M<br>pH=7 | |
|---|---|---|---|
| | | **Temp. = 0°C** | **Temp. = 140°C** |
| $HPO_4^{2-}{}_{(aq)} + 4H_{2(aq)} \rightarrow PH_{3(aq)} + 2H_2O_{(l)} + 2OH^-{}_{(aq)}$ | 225.16 | 410.14 | 538.49 |
| $HPO_4^{2-}{}_{(aq)} + CH_{4(aq)} \rightarrow PH_{3(aq)} + CO_{2(aq)} + 2OH^-{}_{(aq)}$ | 415.32 | 503.97 | 558.16 |
| $HPO_3^{2-}{}_{(aq)} + 3H_{2(aq)} \rightarrow PH_{3(aq)} + H_2O_{(l)} + 2OH^-{}_{(aq)}$ | 130.79 | 303.69 | 355.06 |
| $HPO_3^{2-}{}_{(aq)} + ½H_2O_{(l)} + ¾CH_{4(aq)} \rightarrow PH_{3(aq)} + ¾CO_{2(aq)} + 2OH^-{}_{(aq)}$ | 358.80 | 451.88 | 459.38 |
| $½ Ca_3(PO_4)_{2(s)} + 4H_{2(g)} \rightarrow PH_{3(g)} + 1½Ca(OH)_{2(s)} + H_2O_{(g)}$ | 373.32 | 465.08 | 526.75 |

**2) Reduced atmosphere (high $H_2$ levels)**

| Possible geochemical $PH_3$ formation reaction | ΔG° (kJ/mol) | ΔG (kJ/mol) at plausible geochemical gas concentrations<br>$H_2 = 1$ bar<br>$PH_3 = 10^{-6}$ bar<br>$CH_4 = 10^{-1}$ bar<br>$CO_2 = 0.01$ bar<br>phosphate=$10^{-6}$ M<br>phosphite = $10^{-5}$ M<br>pH=7 | |
|---|---|---|---|
| | | **Temp. = 0°C** | **Temp. = 140°C** |
| $HPO_4^{2-}{}_{(aq)} + 4H_{2(aq)} \rightarrow PH_{3(aq)} + 2H_2O_{(l)} + 2OH^-{}_{(aq)}$ | 225.16 | 289.95 | 356.65 |
| $HPO_4^{2-}{}_{(aq)} + CH_{4(aq)} \rightarrow PH_{3(aq)} + CO_{2(aq)} + 2OH^-{}_{(aq)}$ | 415.32 | 483.07 | 526.54 |
| $HPO_3^{2-}{}_{(aq)} + 3H_{2(aq)} \rightarrow PH_{3(aq)} + H_2O_{(l)} + 2OH^-{}_{(aq)}$ | 130.79 | 204.40 | 204.84 |
| $HPO_3^{2-}{}_{(aq)} + CH_{4(aq)} \rightarrow PH_{3(aq)} + CO_{(aq)} + 2OH^-{}_{(aq)}$ | 358.80 | 427.06 | 421.83 |
| $½ Ca_3(PO_4)_{2(s)} + 4H_{2(g)} \rightarrow PH_{3(g)} + 1½Ca(OH)_{2(s)} + H_2O_{(g)}$ | 373.32 | 339.65 | 337.00 |

**Table C-1:** Reaction pathways and the energy of formation of phosphine from plausible geochemical volatile concentrations. Free energy of formation of $PH_3$ (ΔG) was calculated under geochemical concentration of gases for two different temperature scenarios and two $H_2$ levels (1) oxidized atmosphere (low $H_2$ levels) and (2) reduced atmosphere (high $H_2$ levels). Under all of the tested terrestrial planet conditions the free energy of formation of $PH_3$ (ΔG) is positive, making the geochemical formation of $PH_3$ from phosphate and phosphite an unlikely scenario at circumneutral pH.

## D: Production of Phosphine from Phosphite at Acidic pH

It has been suggested that abiotic disproportionation of phosphites could be a source of environmental phosphine (Roels and Verstraete 2001). Indeed, one common laboratory method of obtaining $PH_3$ is by heating phosphite which disproportionates to phosphine and phosphate (Gokhale *et al.* 2007). This process is only favored at high temperature (T>50 °C, or T>323 K) or acid conditions (pH~<1.3, where $H_3PO_3$ and/or $H_2PO_3^-$ dominate) (Figure D-1). We also note that the production of phosphite *via* the reduction of phosphate with $H_2$ on Earth-analog habitable worlds is highly endergonic and is also unlikely (Table D-1). For a detailed analysis of thermodynamic limitations of $PH_3$ and phosphite production see (Bains *et al.* 2019a).

| Temperature | ΔG (kJ/mol); Oxidizing conditions (as above in Table C-1) | ΔG (kJ/mol); Reducing conditions (as above in Table C-1) |
|---|---|---|
| 0°C/273 K | 38.52 | 17.62 |
| 140°C/413 K | 95.36 | 63.73 |

**Table D-1**: ΔG (kJ/mol) for forming phosphite by reduction of phosphate with $H_2$: $HPO_4^{2-} + H_2 \rightarrow HPO_3^{2-} + H_2O$.

Such conditions are not characteristic of Earth-analog habitable worlds, though they are available at “black smoker” vents (Martin *et al.* 2008), and might be more common on a hot world with more acid oceans, e.g. a world with high $pCO_2$ orbiting close to its parent star. We conclude that disproportionation of phosphite is unlikely to generate significant fluxes of phosphines on Earth-analog habitable worlds, but that worlds which are known to have hot, extremely acid oceans (T>~50 °C / 323 K, pH<~1.3) may be vulnerable to this false positive scenario.

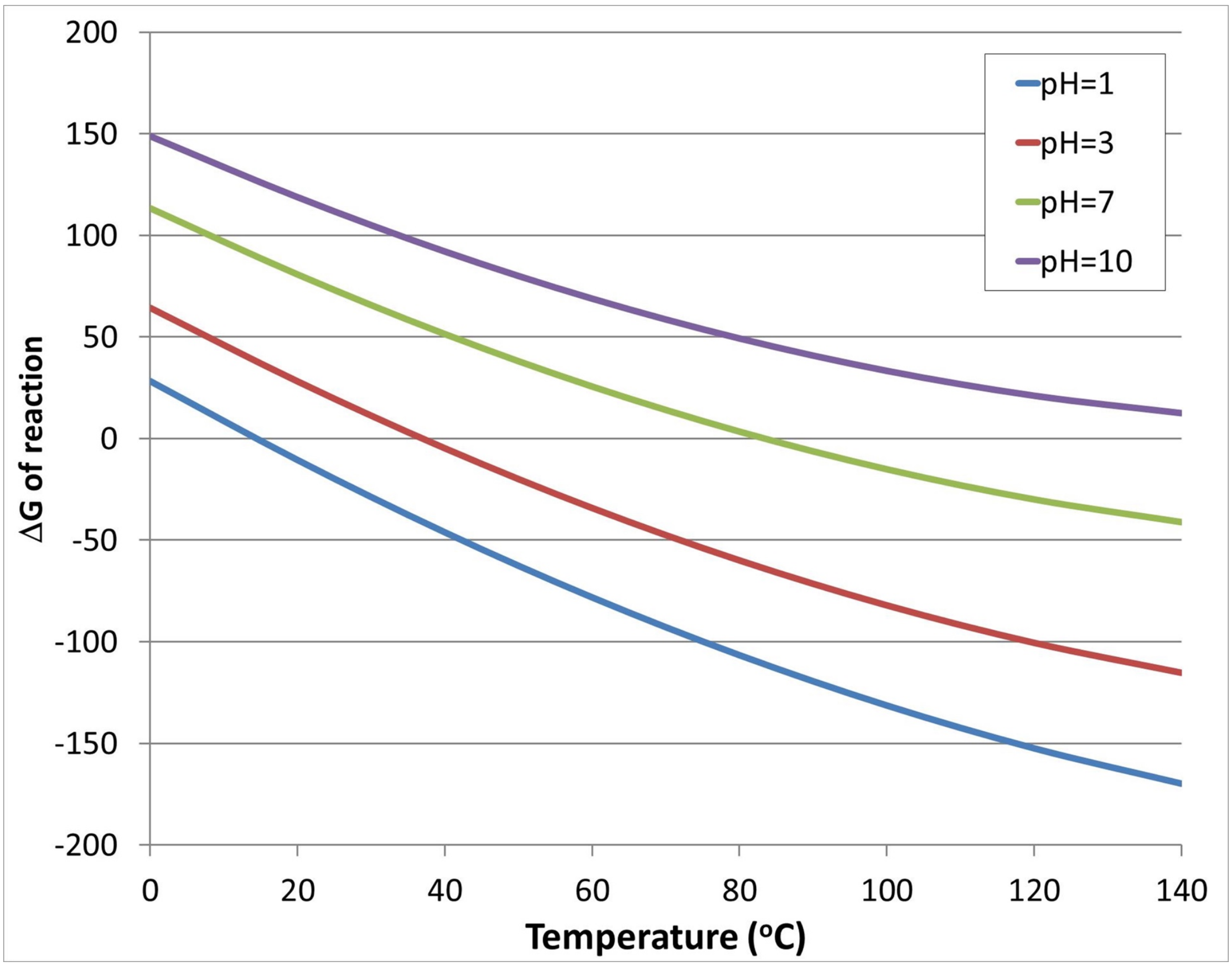

**Figure D-1**: Phosphine production by disproportionation of phosphite is favored under very low pH or high temperature conditions. Y-axis: ΔG° (Gibbs free energy of reaction under standard conditions), in units of kJ/mol. X-axis: temperature (°C). The free energy of reaction for four reactions are shown: disproportionation at very low pH, where phosphite is present as unionized $H_3PO_3$ form; and disproportionation of ionized species present at higher pHs (see Bains *et al.* 2019a for more details on calculation method).

## E: Phosphine in the Context of Terrestrial Biology

We briefly address a few of the common concerns in the association of phosphine with terrestrial biology: the toxicity of $PH_3$; the absence of a known metabolic biosynthetic pathway and enzymatic mechanism for the production of $PH_3$; some context on the ecology of phosphines; and a discussion of the phosphorus cycle.

**Phosphine toxicity**: On modern Earth, phosphine is a rare, toxic gas. It is highly toxic to aerobically metabolizing organisms (Bond and Monro 1967), which is reflected by its wide use as fumigant (e.g. reviewed in (Perkins *et al.* 2015)). The toxicity of $PH_3$ appears to be strictly dependent on aerobic metabolism (e.g. reviewed in (Valmas *et al.* 2008)). The detailed analysis of the toxic effects of $PH_3$ on aerobic organisms is beyond the scope of this paper and is published elsewhere (Bains *et al.* 2019b). However, if $PH_3$ chemistry is indeed selectively incompatible with $O_2$-dependent metabolism of aerobic organisms, as literature suggests, it opens an intriguing possibility that phosphines may safely be produced as secondary metabolites to much greater extent by obligatorily anaerobic life, that does not rely on $O_2$ metabolism (Bains *et al.* 2019b).

**Biosynthetic pathways of phosphine and other volatiles**: The identification of natural products from anaerobic organisms and elucidation of their biosynthetic pathways is notoriously difficult. Culturing of anaerobic organisms is much more laborious task than aerobic ones and identification or purification of anaerobic metabolites is much more complex as well. It requires specialized experimental setup, as often isolated molecules or their biosynthetic precursors get destroyed in our oxygen-rich atmosphere before they can be properly studied. It is therefore not surprising that the exact molecular mechanism of biological phosphine formation in anaerobic environment has eluded discovery for such a long time. Even identification of biosynthetic pathways of natural molecules from aerobic organisms that can be easily cultured in the laboratory conditions can take many decades. For example, the biosynthetic pathway of nucleocidin, an unusual fluorine-containing natural antibiotic produced by a bacterium *Streptomyces calvus* was only recently elucidated, after ~60 years of intensive studies (Petkowski *et al.* 2018).

The economically important production of volatile molecules by fruiting bodies of truffles (*Tuber* sp.), responsible for the unique aroma of the fungus, is another good illustration of the complexity of the biological production of volatile and non-volatile natural metabolites. Despite many decades of studies, and an influx of funding from a multimillion-dollar food industry, the biosynthetic pathways responsible for synthesis of truffle volatiles are not fully understood (Zambonelli *et al.* 2016). The biological production of truffle volatiles, for example thiophene derivatives, is further complicated by the fact that the biosynthetic pathways for molecules responsible for the unique aroma of truffles are likely shared between multiple species of symbiotic microorganisms (bacteria and yeast) living within the fruiting body of the fungus. The metabolic processes of the symbionts and the fungus host collectively allow for the formation of the final volatile product (e.g. thiophene derivatives) (Zambonelli *et al.* 2016). Each of the symbionts, and the host fungus, likely contain only part of the full biosynthetic pathway required for the formation of the final volatile product and each single species is only responsible for the production of few biosynthetic intermediates for the formation of thiophene derivatives (Zambonelli *et al.* 2016).

Such ecological complexity of biosynthetic pathways is a common occurrence. It is quite possible that phosphine production requires a similarly complex ecological scenario, where the full biosynthetic pathway for $PH_3$ synthesis is shared between multiple species of anaerobic microorganisms and only leads to phosphine formation under very specific environmental conditions (Bains *et al.* 2019a). Such complex ecological scenarios further complicate the elucidation of the direct mechanism of biosynthesis. As it is in the case of the complex ecology of biosynthesis of truffle volatiles, the lack of understanding of the details of their biological production does not mean that phosphine or any other natural molecule is not a product of the metabolism of the living organisms.

**Phosphines in the Earth's ecology**: All life on Earth relies on phosphorous-containing compounds in its metabolism. The great majority of biochemicals used by life on Earth are pentavalent phosphorus-containing molecules, predominantly phosphates (Petkowski *et al.* 2019b). At the first glance the trivalent phosphorus compounds appear to be almost completely absent from biology, yet strong circumstantial evidence seems to suggest that anaerobic life on Earth explores chemistry of phosphines beyond simple $PH_3$ production. One isolated study (Davies 2008) reports identification of phospholane volatile in biological samples. Phospholane is a volatile, trivalent phosphorus-containing, five-membered ring saturated hydrocarbon. It is the only molecule reported so far to be isolated from biological samples that contains a bond between trivalent phosphorus and carbon atoms. Phospholane, like phosphine, is a volatile trivalent phosphorus molecule that was identified in European badger fresh scat samples (Davies 2008). It is interesting to note that, so far, all trivalent phosphorus-containing molecules that were isolated from biological samples (i.e. $PH_3$ and phospholane) appear to be produced in strictly anoxic environments and appear to be exclusively associated with anaerobic $O_2$-free dwelling microorganisms, which seems to be in agreement with previous observations that phosphines tend to be poisonous to aerobic life, but non-toxic in anoxic environments.

Both phosphine and phospholane were detected in anaerobic environments which are generally difficult to study. If natural production of trivalent phosphorus-containing natural chemicals is strictly dependent on anoxic environments, it is likely that more such molecules are going to be discovered in the future. Almost all natural compounds reported to be produced by life on Earth were detected or isolated from organisms living in oxygenic environments, as shown by the repositories of natural molecules produced by life (Bains *et al.* 2019b; Petkowski *et al.* 2019a), thus the small number of known natural molecules containing trivalent phosphorus might be the result of sample bias towards $O_2$-rich environment. For a detailed study on the chemistry of phosphine in the context of the terrestrial biology please see (Bains *et al.* 2019b).

**The phosphorus cycle on Earth:** It is generally assumed that life on Earth only consumes phosphorus in its most oxidized form of phosphate, and that phosphates are the only form of phosphorus that is useful for life's cellular metabolism. This assumption is incorrect. Organisms can also use phosphorus in many other forms beyond phosphates, such as organic phosphorus esters, polyphosphates, phosphonates and even reduced phosphorus species like or phosphites. Phosphite and hypophosphite biochemistry is well known (Casida Jr 1960; Figueroa and Coates 2017; Metcalf and Wolfe 1998; Pasek *et al.* 2014; Stone and White 2012) and some microorganisms living in strictly anoxic environments, such as anaerobic sludges in waste-water treatment plants, can fulfill all of their energy requirements by oxidizing phosphite to phosphate (a process called dissimilatory phosphite oxidation) (Figueroa *et al.* 2018; Figueroa and Coates 2017). Studies in recent years established that a complex redox phosphorus cycle (beyond the simple cycling of phosphates) has equally important bioenergetic and ecological consequences as the more recognized microbial nitrogen, carbon or sulfur cycles (e.g., (Karl 2014)).

Marine organisms can reduce a significant fraction (up to 15%) of the total phosphates taken up from the surrounding environment (Van Mooy *et al.* 2015). The reduced phosphorus species (e.g., in the form of the soluble phosphite) are then rapidly released to the surrounding sea water (Van Mooy *et al.* 2015). The authors argue that the total amount of phosphorus cycled through such complex redox reactions could exceed other oceanic sources of phosphorus (e.g., direct continental and atmospheric sources) (Van Mooy *et al.* 2015). It is unknown why marine planktonic organisms engage in such an energetically costly endeavor. Production and rapid release of large quantities of reduced phosphorus species is unexpected, especially in the open oceanic waters where phosphorus is a limiting nutrient. The production of reduced phosphorus species is, of course, dependent on the abundance of phosphorus, but appears to be independent from inorganic or organic phosphorus concentrations, and independent from the overall phosphate turnover time (Van Mooy *et al.* 2015). The synthesis and rapid export of soluble reduced phosphorus species could stem from the general cooperative behavior of planktonic organisms in the open ocean. One of the more studied producers of reduced phosphorus species is the cyanobacteria *Trichodesmium.* The colonies of *Trichodesmium* are also inhabited by many species of epibiotic and symbiotic bacteria that appear to cooperate in the cycling of phosphorus (Hmelo *et al.* 2012; Van Mooy *et al.* 2012). Phosphonates and reduced phosphorus compounds (i.e. phosphites) may be part of a barter economy where phosphorus is exchanged for other nutrients between *Trichodesmium* and other species living in the cooperative colonies. Such examples of biological importance of rare redox phosphorus species clearly show that reduced phosphorus species play an important albeit under-appreciated role in the global biogeochemical cycle of phosphorus.

The exact place and role of phosphine in the global phosphorus cycle is not yet fully known. It is, however, likely that, similarly to other reduced phosphorus species, $PH_3$ also has an important role in the global cycling of this essential element.